%% file: confid.tex
\documentclass[10pt,journal,compsoc]{IEEEtran}
\usepackage{url}
\usepackage{epsfig,tabularx,amssymb,amsmath,amssymb,dsfont,subfigure,multirow,
algorithmic,graphicx}
\usepackage{subfigure}
\usepackage{multirow}
\usepackage{times,epsfig}
\usepackage{dblfloatfix}
\usepackage[vlined,ruled,linesnumbered]{algorithm2e}
\usepackage{epsfig,tabularx,amssymb,amsmath,subfigure,multirow,
algorithmic,graphicx}
\usepackage{color,graphicx}
\usepackage{subfigure,multirow}
\usepackage{times}
\usepackage{amsmath}
\usepackage{epsfig}
\usepackage{ctable}
\usepackage{threeparttable}
\usepackage{enumerate}
\usepackage{amsmath}
\usepackage{amsfonts}
\usepackage{color}
\usepackage{multirow}
\usepackage{graphicx}
\usepackage{balance}
\usepackage{paralist}
\usepackage{CJKutf8}
\usepackage{url}
\usepackage{amssymb}

\usepackage{multirow}

\newcommand{\nop}[1]{}
\newtheorem{example}{Example}

\newtheorem{problem}{Problem~}

\newtheorem{definition}{Definition}

\newtheorem{Candidate Grid Matrix}[Grid]{Definition}
\newtheorem{Location point}[Grid]{Definition}

\newtheorem{Repair Observation Probability}[Grid]{Definition}
\newtheorem{Transmission Probability}[Grid]{Definition}

\begin{document}
\title{Context-aware Telco Outdoor Localization}

\author{Yige Zhang, Weixiong Rao, Mingxuan Yuan, Jia Zeng and Pan Hui
\IEEEcompsocitemizethanks{
\IEEEcompsocthanksitem Yige Zhang and Weixiong Rao are with School of Software Engineering, Tongji University, Shanghai, China. \protect
E-mail: \{yigezhang, wxrao\}@tongji.edu.cn
\IEEEcompsocthanksitem Mingxuan Yuan and Jia Zeng are with Huawei Noahs Ark Lab, Hong Kong. \protect
E-mail: \{mingxuan.yuan, jia.zeng\}@huawei.com
\IEEEcompsocthanksitem Pan Hui is with Department of Computer Science and Engineering, Hong Kong University of Science and Technology, and Department of Computer Science, University of Helsinki.  \protect
E-mail: panhui@cse.ust.hk
}
\thanks{}}

\maketitle
\input{00-abstract}

\input{01-intro}

\input{02-problem}

\input{03-hmm}

\input{04-repair}

\input{05-evaluate}

\input{06-conclusion}

\bibliographystyle{abbrv}
\bibliography{localization}

\scriptsize
\begin{IEEEbiography}[{\includegraphics[width=.8in,height=1.0in,clip, keepaspectratio] {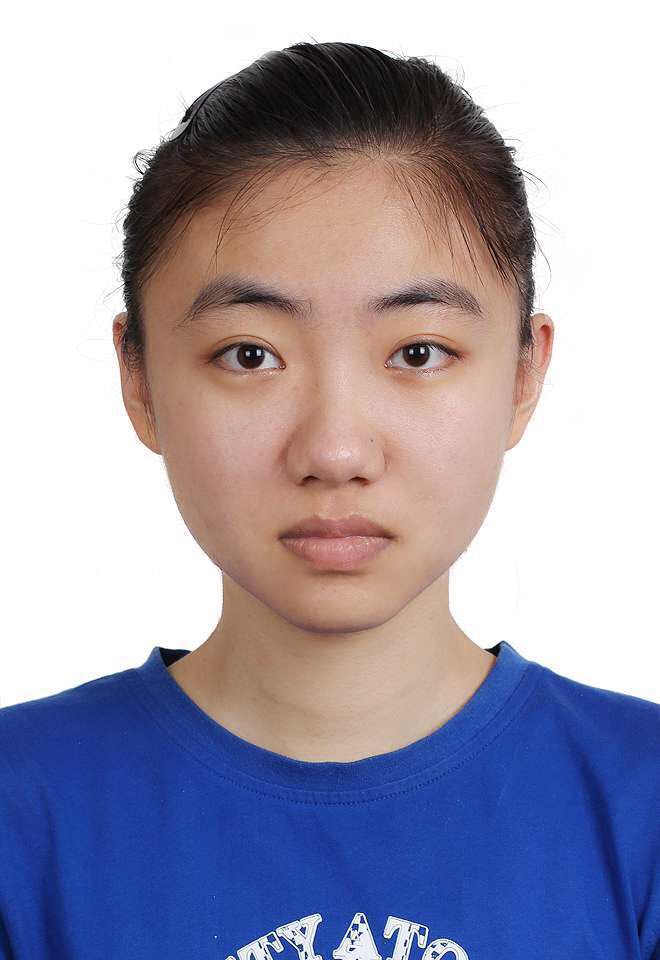}}]
{Yige Zhang} is a PhD student in School of Software Engineering, Tongji University, China, and received the B.Sc degree from Tongji University in 2016. Her research interests focus on mobile computing and data mining.
\end{IEEEbiography}\vspace{-4ex}
\begin{IEEEbiography}[{\includegraphics[width=.8in,height=1.0in,clip, keepaspectratio] {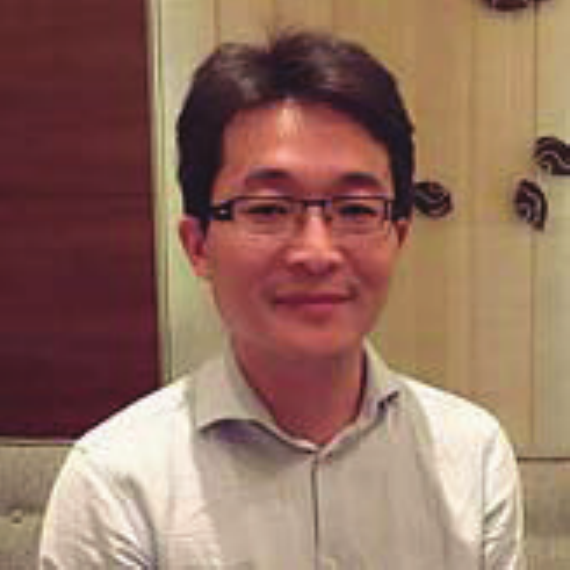}}]
 {Weixiong Rao} received his Ph.D degree from The Chinese University of Hong Kong in 2009. After that, he worked for Hong Kong University of Science and Technology (2010), University of Helsinki (2011-2012), and University of Cambridge Computer Laboratory Systems Research Group (2013) as Post-Doctor. He now is a Professor in School of Software Engineering, Tongji University, China. His research interests include mobile computing and spatiotemporal data science.
\end{IEEEbiography}
\vspace{-4ex}
\begin{IEEEbiography}[{\includegraphics[width=.8in,height=1.0in,clip, keepaspectratio] {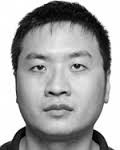}}]
{Mingxuan Yuan} received the PhD degree from Hong Kong University of Science and Technology. He is a researcher with Noah's Ark Lab, Huawei. His main research interests include spatiotemporal data management/mining, telco (telecommunication) big data management/mining, telco big data privacy, and visualization. He is a
member of the IEEE.
\end{IEEEbiography}\vspace{-4ex}
\begin{IEEEbiography}[{\includegraphics[width=.8in,height=1.0in,clip, keepaspectratio] {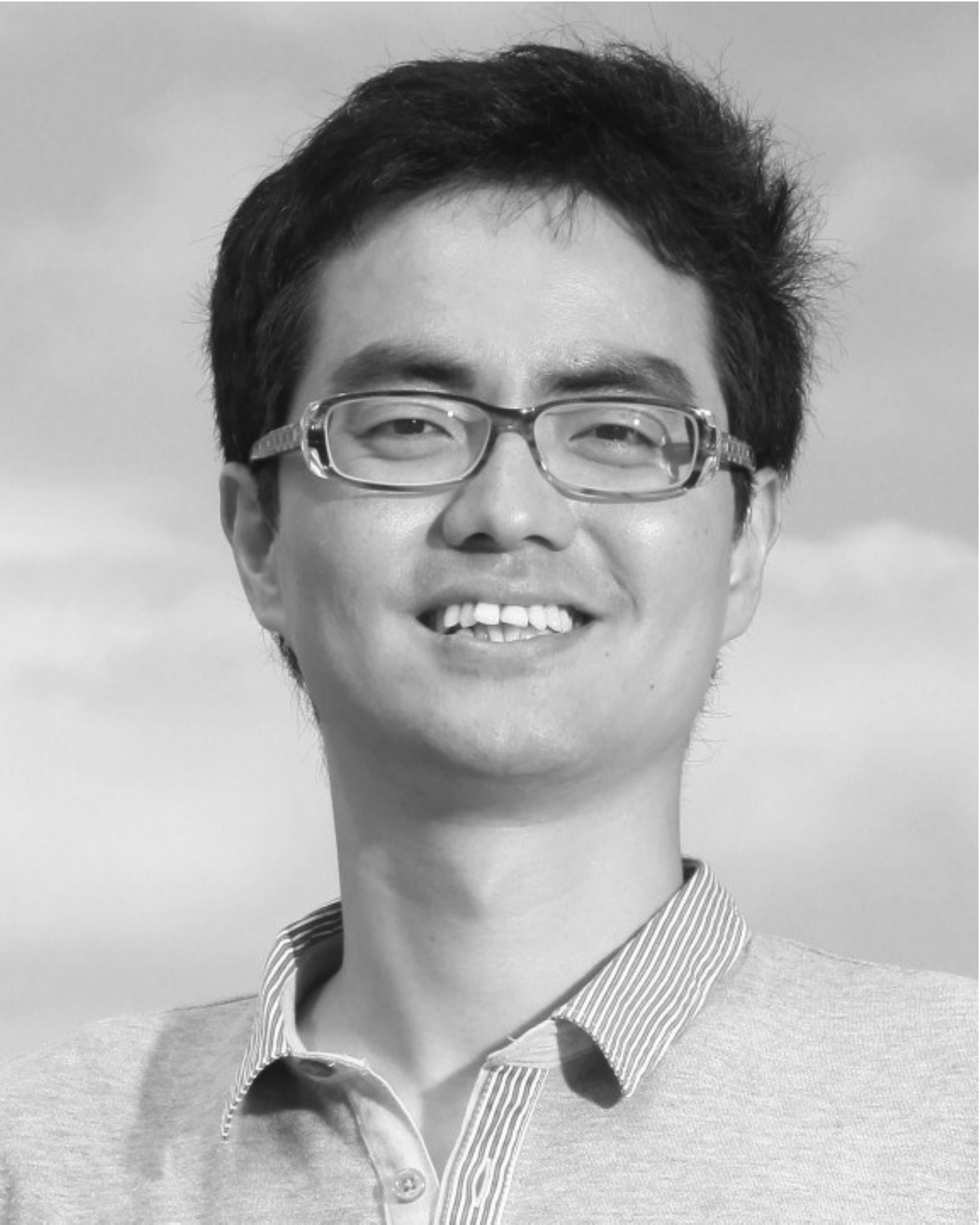}}]
{Jia Zeng} is a principal researcher at Noah's Ark Lab, Huawei. He obtained his Ph.D. degree from City University of Hong Kong. His main research interests include scalable machine learning algorithms, telco big data analytics and enterprise intelligence (supply chain, enterprise finance, IT supporting system and etc.).
\end{IEEEbiography}\vspace{-4ex}
\begin{IEEEbiography}[{\includegraphics[width=.8in,height=1.0in,clip, keepaspectratio] {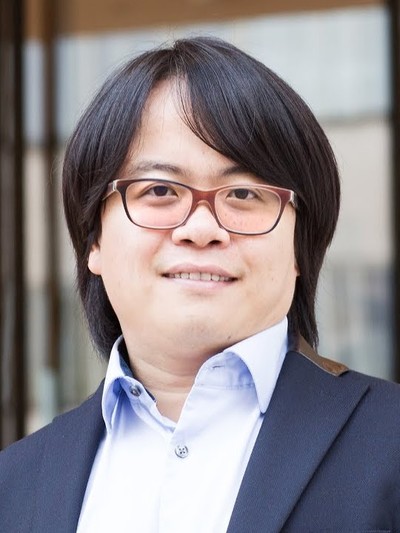}}]
{Pan Hui} received his PhD from the Computer Laboratory at University of Cambridge, and both his Bachelor and MPhil degrees from the University of Hong Kong. He is the Nokia Chair Professor in Data Science and Professor of Computer Science at the University of Helsinki. He is also the director of the HKUST-DT System and Media Lab at the Hong Kong University of Science and Technology. He has published more than 300 research papers and with over 18,000 citations. He has 30 granted and filed European and US patents. He has founded and chaired several IEEE/ACM conferences/workshops, and has served as track chair, senior program committee member, organising committee member, and program committee member of numerous top conferences including ACM WWW, ACM SIGCOMM, ACM Mobisys, ACM MobiCom, ACM CoNext, IEEE Infocom, IEEE ICNP, IEEE ICDCS, IJCAI, AAAI, and ICWSM. He is an Associate Editor for IEEE Transactions on Mobile Computing (since 2014) and the Springer journal of Computational Social Networks. He has also served as Associate Editor for IEEE Transactions on Cloud Computing (2014 - 2018) and guest editor for various journals including IEEE Journal on Selected Areas in Communications (JSAC), IEEE Transactions on Secure and Dependable Computing, IEEE Communications Magazine, and ACM Transactions on Multimedia Computing, Communications, and Applications. He is an ACM Distinguished Scientist, an IEEE Fellow, and a member of Academia Europaea.

\end{IEEEbiography}
\end{document}

%% file: 00-abstract.tex
\begin{abstract}
Recent years have witnessed the fast growth in telecommunication (Telco) techniques from 2G to upcoming 5G. Precise outdoor localization is important for Telco operators to manage, operate and optimize Telco networks. Differing from GPS, Telco localization is a technique employed by Telco operators to localize outdoor mobile devices by using measurement report (MR) data. When given MR samples containing noisy signals (e.g., caused by  Telco signal interference and attenuation), Telco localization often suffers from high errors. To this end, the main focus of this paper is how to improve Telco localization accuracy via the algorithms to detect and repair outlier positions with high errors. Specifically, we propose a context-aware Telco localization technique, namely \textsf{RLoc}, which consists of three main components: a machine-learning-based localization algorithm, a detection algorithm to find flawed samples, and a repair algorithm to replace outlier localization results by better ones (ideally ground truth positions). Unlike most existing works to detect and repair every flawed MR sample independently, we instead take into account spatio-temporal locality of MR locations and exploit trajectory context to detect and repair flawed positions. Our experiments on the real MR data sets from 2G GSM and 4G LTE Telco networks verify that our work \textsf{RLoc} can greatly improve Telco location accuracy. For example, \textsf{RLoc} on a large 4G MR data set can achieve 32.2 meters of median errors, around 17.4\% better than state-of-the-art.   \end{abstract}

%% file: 01-intro.tex
\section{Introduction}

Outdoor localization systems have gained focus recently due to the remarkable proliferation of telecommunication (Telco) networks (from 2G to upcoming 5G networks) and sensor-rich smart mobile devices. These systems span different application domains, such as navigation systems, location-based advertisements, social networks and resource allocation in wireless networks \cite{DBLP:journals/comsur/LaoudiasMKLWF18}. In particular, Telco operators have strong interest in localization technology due to their needs for automated network management, operation and optimization. Specifically, \emph{location information} of mobile devices is important for Telco operators to \emph{1}) identify location hotspots for capacity planning, \emph{2}) identify gaps in radio frequency coverage, \emph{3}) troubleshoot network anomalies, and \emph{4}) locate users in emergency situations (E911) \cite{MargoliesBBDJUV17}.

Differing from GPS, Telco localization is a technique employed by Telco operators to localize outdoor mobile devices by using measurement report (MR) data. MR data mainly contain the connection information, such as signal strength, between mobile devices and nearby base stations. Telco operators exploit a backend localization algorithm on MR data to infer the locations of mobile devices. Due to the rich commercial opportunities of the inferred locations, Telco localization has recently attracted intensive research interests in both academia \cite{aly2013dejavu,vaghefi2011rss,IbrahimY12} and Telco industries \cite{MargoliesBBDJUV17,RayDM16,ZhuLYZZGDRZ16,ZangBB10}. %including AT\&T, Huawei and Sprint 

Unfortunately, the design of an accurate Telco localization algorithm is challenging. For example, high buildings in urban cities often cause Telco signal interference and attenuation. Mobile devices located in those areas with high buildings often generate noisy MR samples containing unstable signal strength. When given such MR samples, Telco localization cannot achieve high accuracy. Though the recently popular data-driven localization \cite{IbrahimY12,ZhuLYZZGDRZ16,ChakrabortyOD15} leverages those MR samples tagged by GPS coordinates to train a machine-learning-based localization model, the localization accuracy is around $80$ meters in terms of median errors \cite{ZhuLYZZGDRZ16}, leading to little chance of achieving GPS-like performance \cite{ChakrabortyOD15}.

In this paper, we propose a context-aware Telco localization technique, namely \textsf{RLoc}, in order to achieve high localization accuracy. Our work is motivated by the following observation. For those MR samples containing noisy signals, their predicted positions are typically with high errors and significantly degrade overall localization accuracy. For simplicity, the samples leading to high errors are called \emph{flawed samples}, and corresponding predicted locations with high errors are called \emph{flawed locations}. To this end, the main focus of this paper is how to improve Telco localization accuracy via the algorithm to detect flawed samples and repair flawed positions. That is, we would like to first detect flawed MR samples. If the associated flawed positions can be repaired by highly precise ones (ideally ground truth positions), we then have chance to achieve much lower localization errors. Nevertheless, most existing works detect each individual flawed sample and then repair the corresponding flawed position~\cite{DBLP:conf/kdd/GaoLFWSH10,DBLP:journals/tkdd/LiuTZ12,ZhangRYZY17,DBLP:conf/ijcai/LiDHL17} and do not take into account contextual knowledge of neighbouring MR positions. Unlike these works, we consider that a sequence of MR positions exhibits spatio-temporal locality and contributes to a trajectory of positions. For example, when a mobile device is moving around high buildings and suffers from Telco signal interference, we assume that a sequence of generated MR samples is all flawed. By exploiting the spatio-temporal context in the trajectory of MR positions, we design the sequence-based detect and repair approach for much lower errors. As a summary, we make the following contributions.

\begin{itemize}

\item \emph{Confidence-based detection algorithm}: Based on the physical distance between predicted position and real ones, we define a \emph{confidence level} for an MR sample to determine whether or not the sample is flawed. Beyond that, we are interested in the confidence levels of a sequence of MR samples. Thus, we propose a dual-stage adaptive Hidden Markov Model, called DA-HMM, to predict a corresponding sequence of confidence levels. By introducing the adaptive state transition probability and adaptive mission probability, DA-HMM can process real world MR sequences (which exhibit uneven timestamp intervals among neighbouring MR samples), and thus lead to better performance than traditional HMM models.
    
\item \emph{Joint probability-based repair algorithm}: Still when given a sequence of flawed MR samples, we are interested in not only the goodness of a certain candidate position to repair an individual flawed position, but also the transition possibility from the previous position to the next one. Thus, we define the \emph{joint probability} of an entire path to connect candidate positions. Among all possible paths of candidate positions, we design a dynamic planning algorithm to select the best one to repair the entire sequence of flawed positions.
\item \emph{Extensive Performance Validation}: Our experiments on the real MR data sets from 2G GSM and 4G LTE Telco networks verify that our work \textsf{RLoc} greatly improves Telco location accuracy. For example, \textsf{RLoc} on a large 4G MR data set can achieve 32.2 meters of median error. Such numbers indicate that \textsf{RLoc} achieves comparable accuracy as GPS.
\end{itemize}

The rest of this paper is organized as follows. Section \ref{s:problem} first reviews the background and related work. Section \ref{sec:prob} then formulates the problem definition and highlights the solution. Next, Sections \ref{sec:hmm} and \ref{sec:rrepair} describe the detection and repair algorithms, respectively. After that, Section \ref{s:evaluation} evaluates our work. Section \ref{s:conclusion} finally concludes the paper. Table \ref{tab:symbols} summarizes the mainly used terms/symbols and associated meanings.

\begin{table}[!htb]%\hspace*{-4ex}
 \caption{Used Terms and Associated Meanings}
 \label{tab:symbols}%\vspace{-3ex}
\centering\scriptsize
\begin{tabular}{|l|l|}\hline
\textbf{Term/Symbol} & \textbf{Meaning}\\
\hline\hline
MR & Measurement Report \\%\hline
RSSI & Radio Signal Strength Index  \\%\hline
Telco & Telecommunication \\%\hline
HMM & Hidden Markov Model \\%\hline
DA-HMM & A Dual-stage Adaptive Hidden Markov Model \\\hline%\hline
$r$&MR sample\\%\hline
$L_p(r)$&Predicted location of MR sample $r$\\%\hline
$L_t(r)$ &Ground truth location of MR sample $r$\\
$R=\{r_1,...,r_{|S|}\}$ & a sequence of $|R|$ MR samples\\
\hline%\hline
$\mathcal{L}$& Telco localization model\\%\hline
$\mathcal{C}$ & Confidence model\\\hline%\hline

$\mathbb{D}$& Original Training dataset  with $\mathbb{D}$ = $\mathbb{D}_L \cup \mathbb{D}_C$\\%\hline
$\mathbb{D}_L$& Training subset for localization $\mathcal{L}$\\%\hline
$\mathbb{D}_C$& Training subset for confidence model $\mathcal{C}$ \\\hline%\hline

$\mathfrak{D}$& Testing dataset with $\mathfrak{D} = \mathfrak{D}^- \cup \mathfrak{D}^+$\\%\hline
$\mathfrak{D}^-$& Flawed Testing datasets\\%\hline
$\mathfrak{D}^+$& Non-Flawed Testing datasets\\\hline%\hline

$A=\{a_{i,j}\}$ & State transition probability in HMM \\%\hline
$B=\{b_j(k)\}$ & Emission probability in HMM \\%\hline
$a^\Delta$ & Adaptive state transition prob. by time interval $\Delta$\\%\hline
$b^{\gamma}$ & Adaptive emission prob. by sample size $\gamma$ \\%\hline
$v_k=\{v_k^{bs}, v_k^{ss}\}$   & \begin{tabular}[c]{@{}l@{}} MR observation $v_k$ with a pair of base stations $v_k^{bs}$ \\and Telco signal strength $v_k^{ss}$\\\end{tabular} \\\hline
  \end{tabular}%\vspace{-3ex}

\end{table}

%% file: 02-problem.tex
\section{Background and Related Work}\label{s:problem}

\subsection{Background of MR Data}\label{sec:bkmrdata}

A Measurement Report (MR) sample maintains the connection state of a certain mobile device in a Telco network, including a unique ID (IMSI: International Mobile Subscriber Identity), connection time stamp (MRTime), up to 7 nearby base stations (RNCID and CellID) \cite{Rizk2019CellinDeep}, and corresponding signal measurements such as AsuLevel, SignalLevel and RSSI. Table \ref{tab:mr} gives an example 2G GSM MR sample collected by an Android device. AsuLevel, i.e., Arbitrary Strength Unit Level, is an integer proportional to the received signal strength measured by the mobile device. SignalLevel indicates the power ratio (typically logarithm value) of the output signal of the device and the input signal. RSSI denotes a radio signal strength indicator. Among the up to 7 base stations, one of them is selected as the primary serving station to provide communication and data services for mobile devices.

\begin{table}[hbt]
\caption{An Example of 2G GSM MR Record Collected by an Android Device.}\label{tab:mr}
\scriptsize%\hspace{-3ex}
\centering
\begin{tabular}{|lllll|}
\hline
$\textbf{MRTime}$ ***&	$\textbf{IMSI}$ ***&	$\textbf{SRNC\_ID}$ 6188&	$\textbf{BestCellID}$ 26050&	$\textbf{\#\_BS}$	7%\\
% 2018/4/23 9:20	&xxx&	6188 &	26050	& 7
\\\hline
$\emph{\textbf{RNCID\_1}}$ 6188&	$\emph{\textbf{CellID\_1}}$ 26050&	$\emph{\textbf{AsuLevel\_1}}$ 18&	$\emph{\textbf{SignalLevel\_1}}$ 4&	$\emph{\textbf{RSSI\_1}}$ -77\\
%6188&26050 &	18 &4	&-77 \\
\hline
$\emph{\textbf{RNCID\_2}}$ 6188&	$\emph{\textbf{CellID\_2}}$ 27394&	$\emph{\textbf{AsuLevel\_2}}$ 16&	$\emph{\textbf{SignalLevel\_2}}$ 4&	$\emph{\textbf{RSSI\_2}}$ -81\\
%& &	 &	&\\%
\hline
$\emph{\textbf{RNCID\_3}}$ 6188&	$\emph{\textbf{CellID\_3}}$ 27377&	$\emph{\textbf{AsuLevel\_3}}$ 15&	$\emph{\textbf{SignalLevel\_3}}$ 4&	$\emph{\textbf{RSSI\_3}}$ -83\\
 %& &	 &	& \\%
 \hline
$\emph{\textbf{RNCID\_4}}$ 6188&	$\emph{\textbf{CellID\_4}}$ 27378&	$\emph{\textbf{AsuLevel\_4}}$ 15&	$\emph{\textbf{SignalLevel\_4}}$ 4&	$\emph{\textbf{RSSI\_4}}$ -83\\
%6188 & 27378 &	15 &4	&-83\\%
\hline
$\emph{\textbf{RNCID\_5}}$ 6182&	$\emph{\textbf{CellID\_5}}$ 41139&	$\emph{\textbf{AsuLevel\_5}}$ 16&	$\emph{\textbf{SignalLevel\_5}}$ 4 &	$\emph{\textbf{RSSI\_5}}$ -89\\
%6182 & 41139&	16 &4	&-89 \\%
\hline
$\emph{\textbf{RNCID\_6}}$ 6188&	$\emph{\textbf{CellID\_6}}$ 27393&	$\emph{\textbf{AsuLevel\_6}}$ 9&	$\emph{\textbf{SignalLevel\_6}}$ 3&	$\emph{\textbf{RSSI\_6}}$ -95\\
% 6188 & 27393&   9&3	    &-95\\%
\hline
$\emph{\textbf{RNCID\_7}}$ 6182&	$\emph{\textbf{CellID\_7}}$ 26051&	$\emph{\textbf{AsuLevel\_7}}$ 9&	$\emph{\textbf{SignalLevel\_7}}$ 3&	$\emph{\textbf{RSSI\_7}}$ -95\\
%6182 & 26051&  9&3	   &-95\\
\hline
\end{tabular}

\end{table}

Generally, we can collect MR samples from two typical data sources: \emph{1)} the data collected from client side and \emph{2)} the one from backend Telco operators.
MR samples, no matter generated by either 4G LTE networks or from 2G GSM networks, follow the same data format if they are collected by Android APIs. Nevertheless, the data format of MR samples collected by backend Telco operators may differ from the one by frontend Android APIs (The detail refers to \cite{DBLP:conf/mdm/HuangRZLYZY17}). All these MR samples provide useful data collection sources. Due to the difference between MR data formats by frontend Android devices and backend Telco operators, we use those MR feature items, e.g., RSSI, that appear within all data sets without loss of generality.

\subsection{Related Work on Telco Localization}
Depending upon location results, we category literature works into single-point-based and sequence-based localization. The former works independently process every MR sample to localize an outdoor mobile device, and the latter ones frequently take as input a sequence of MR samples and then leverage the underlying spatio-temporal locality of such MR samples to generate a trajectory of predicted locations.

\subsubsection{ Single-point-based Telco localization}
In terms of single-point-based localization, we classify  literature works into two categories. Firstly, the \emph{distance-based approaches} \cite{Caffery1998Overview} typically use point-to-point absolute distances or angles to localize mobile devices. Geometric techniques are used to triangulate the locations of mobile devices from 3 or more channel measurements of nearby access points, e.g., signal strength and angle-of-arrival \cite{gezici2008survey,cong2002hybrid}. To localize users with information regarding only one base station in a cellular network, the previous work \cite{ZangBB10} proposed a Bayesian inference-based localization approach by incorporating additional measurements (such as round-trip-time, signal to noise and interference ratio: SINR) with the knowledge of network layout. However, these methods usually suffer from low localization accuracy due to multi-path propagation, non-line-of-light propagation and multiple access interference.

Secondly, \emph{machine learning approaches} \cite{HanK2000} either construct a fingerprinting database or train a learning model such as Random Forest (RaF) \cite{ZhuLYZZGDRZ16} and deep neural network (DNN) \cite{ZhangRZYZ19}, from training MR samples to the associated positions. As baseline machine learning approaches, \emph{fingerprinting methods} \cite{IbrahimY12,PaekKSG11,RayDM16,MargoliesBBDJUV17} in general have better accuracy than the aforementioned distance-based approaches, and their average errors are 100 -- 200 meters. The classic work CellSense \cite{IbrahimY12} first divides an area of interest into smaller grid cells and constructs a fingerprint database to store the mapping function between RSSI features to the corresponding grid cells. When given a query (i.e., an input RSSI feature), the online prediction phase searches the fingerprint database to find the $K$ nearest neighbors (KNN) and returns an average weighted location of the $K$ neighbors. A better CellSense-hybrid technique consists of the rough and refinement estimation phases. In a recent work \cite{MargoliesBBDJUV17}, the AT\&T researchers developed an improved fingerprinting-based outdoor localization system NBL, by assuming a Gaussian distribution of signal strength within each divided grid, and it computes the predicted location by using either Maximum Likelihood Estimation (MLE) or Weighted Average (WA).
Unlike the above fingerprinting methods, the \emph{learning-based localization} trains either a multi-classification or a regression model depending upon the representation of MR positions, e.g., spatial grid cells or numeric GPS coordinates. For example, the previous work \cite{ZhuLYZZGDRZ16} proposed a regression model implemented by a two-layer context-aware coarse-to-fine Random Forests (CCR). In addition, a previous work \cite{ChakrabortyOD15} exploits semi-supervised and unsupervised machine learning techniques to reduce the cost of collecting labelled training samples meanwhile without compromising the accuracy of localization.

\emph{Comparison}: we note that distance-based approaches do not require an offline phase to either construct the fingerprinting database or to train the machine learning models, and instead leverage radio signals to localize mobile devices via a Telco signal propagation model. Machine learning-based approaches require sufficient training samples during the offline phase, leading to much higher localization precision than distance-based approaches. These machine learning approaches are frequently called \emph{data-driven localization}.

\subsubsection{Sequence-based Telco localization} Unlike single-point-based localization, sequence-based localization approaches \cite{RayDM16,DBLP:conf/gis/GambsKC10,Killijian2012Next,Asahara2011Pedestrian,ThiagarajanRBMG10,Arthi2010Localization,Ergen2014RSSI,Ibrahim2010A,DBLP:conf/IEEEcit/NiWTYS17,8939387} first group MR samples by IMSI and then sort the grouped MR samples by time stamps, generating the \emph{sequences of neighbouring MR samples}. By mapping the sequential MR samples into trajectories of locations, these approaches exploit contextual information, e.g., spatio-temporal locality, to achieve more accurate localization than single point-based methods.

To enable the sequence-based localization, various HMM-based localization algorithms have been developed, such as \cite{RayDM16,ThiagarajanRBMG10,Arthi2010Localization,Ergen2014RSSI,Ibrahim2010A,8939387}. For example, the previous work \cite{ThiagarajanRBMG10} explored a two-layer-HMM model: {Grid Sequencing} maintains the mapping from a series of GSM fingerprints to a sequence of spatial grid cells, and {Segment Matching} the mapping from the sequence of grid cells to a road map. The previous work CAPS (Cell-ID Aided Positioning System) \cite{PaekKSG11} uses a cell-ID sequence matching technique to estimate current position based on the history of cell-ID and GPS position sequences that match the current cell-ID sequence. This approach essentially identifies user position on a route that he or she ever passed in the past. The work \cite{RayDM16} utilized HMM and particle filtering to localize a sequence of MR samples. A recent work \cite{DBLP:conf/IEEEcit/NiWTYS17} localized mobile devices by using 4G Long-term evolution (LTE) TA (Timing Advance) and RSRP (Reference Signal Receiving Power), by incorporating route constraint (e.g., road networks) for the motion of vehicles into HMM.

In general, our work belongs to the sequence approach. Nevertheless, there exists some significant difference between the previous sequence approaches above and ours. The previous works above such as \cite{RayDM16, DBLP:conf/IEEEcit/NiWTYS17} take the locations of mobile devices (e.g., the divided grid cells in physical space either with road constraints or not) as HMM states. One issue of using such states is that the amount of states is tremendously large and the transition probability is rather sparse and inaccurate with insufficient MR samples. In contrast, we take the developed confidence levels (with the binary values either 0 or 1) as the states. The key point is that even with scarce training samples used for HMM, we still have chance to develop a much accurate localization model. In addition, a recent work \cite{8939387} requires the third-party historical position trajectory database as the prior of HMM. In case that the positions to predict do not follow the similar distribution as the third-party database, the work \cite{8939387} may not work well.

Finally, though our work and CAPS \cite{PaekKSG11} share some commonality in terms of the sequence-based techniques, there exists some significant difference between the two works. Firstly, beyond cell-IDs, our work further leverages signal measurements for more precise localization. Secondly, our work leverages the sequence-based post-processing technique to detect and repair outlier positions and instead CAPS targets the sequence-based localization. In some sense, the proposed post-processing technique can improve the positions generated by CAPS.  Finally, in terms of the sequence-based algorithm, we mainly exploit the improved HMM-based detection and a dynamic-programming (DP)-based repair algorithm. Instead, CAPS, among a historical Cell-ID sequence database, finds out the sequences that are similar to the currently observed sequence via a sequence matching algorithm, e.g., Smith-Waterman.

\subsection{Related Work on Outlier Detection and Repair}\label{sec:relate}

\textbf{Outlier detection}: To perform data repair, we first need to detect flawed MR samples or outliers. In general, outlier detection methods include statistic approaches, proximity-based, clustering-based and classification-based approaches \cite{HodgeA04,HanK2000}. The first three approaches frequently assume that normal objects either \emph{1}) follow a statistical/stochastic model (e.g., Gaussian distribution), or \emph{2}) are close with the nearest neighbors in feature space, or \emph{3}) belong to large and dense clusters, respectively; and otherwise the remaining objects then become outliers. Differing from the three approaches above, classification-based approaches train a classification model (with two classes) to distinguish normal objects from outlier ones.

We detect flawed MR samples differs from the approaches above. The three approaches above all perform outlier detection directly on MR samples or associated features. Instead, we do not detect whether or not a certain MR sample $r$ is flawed, and instead detect 
whether or not the prediction result of $r$ is an outlier. It makes sense because we are interested in outlier locations, instead of  outlier MR samples or MR features.

\textbf{Data repair}: Once outlier objects are detected, the simplest way is to discard them. Instead, data repair techniques replace outlier objects with either existing normal objects or newly created objects. The key of data repair is a \emph{minimal repair principle}, i.e., to minimize the distortion between original data and repaired data based on some semantic constraints and/or rules. In a recent work targeting on GPS points, Song etc. \cite{SongLZ15} proposed to repair a noise GPS point by an existing GPS point within a cluster, such that data repair and clustering co-occur together (instead of separating data repair from data clustering) with the objective to minimize repair cost. The previous work \cite{AliPM12} targeted the data cleaning in Wireless Sensor Network (WSN) and establishes belief on spatially related nodes to identify potential nodes that can contribute to data cleaning. In addition, to repair a spatial-temporal database, the previous works \cite{EmrichKMRTZ13,MauderREZRTT15} defined spatial-temporal constraints (such as an object must not enter a specified area on Sunday 2am and 5am) and the repair objective is to minimize the change between initial database and repaired database.

Our work differs from the works above. \emph{1}) We do not repair flawed MR samples directly, and instead repair the associated locations. In this way, we have change to optimize the accuracy of the proposed localization algorithm. \emph{2}) Unlike the work \cite{AliPM12}, we do not evaluate the confidence of mobile devices, but the confidence of predicted locations. It makes sense because flawed MR samples are typically caused by high buildings in urban cites. \emph{3}) Finally, the traditional data repair approaches frequently exploited integrity constraints. Without the predefined constraints, such approaches do not work very well \cite{SongZW16}. In our case, it is non-trivial to find data repair constraints in Telco localization. We therefore employ machine learning algorithms to repair prediction result, but not MR samples themselves.

\section{Solution Overview}\label{sec:prob}

\subsection{Problem Definition}\label{s:prob}
Consider that we train a localization model $\mathcal{L}$ from a training dataset $\mathbb{D}$, and then predict the locations of MR samples in a testing MR dataset $\mathfrak{D}$. We are interested in the quality of these predicted positions. Specifically, consider that the localization model $\mathcal{L}$ generates a trajectory of positions for an input sequence of testing MR samples in $\mathfrak{D}$. For each testing sample $r\in \mathfrak{D}$, $\mathcal{L}$ predicts a location $L_p(r)$. Denote the ground truth position of $r$ by $L_t(r)$. If a mobile device located at the true position $L_t(r)$ suffers from Telco signal interference (e.g., caused by nearby high buildings), $L_p(r)$ could significantly differ from $L_t(r)$ and the Euclidean distance between $L_p(r)$ and $L_t(r)$, denoted by $||L_p(r)-L_t(r)||$, is non-trivial. Here, the challenge is that, no matter which and how a certain  algorithm is applied to train the localization model $\mathcal{L}$, the distance $||L_p(r)-L_t(r)||$ (a.k.a localization error) is still high. Thus, we would like to \emph{detect} those samples $r$ suffering from high errors, and then \emph{repair} the predicted locations $L_p(r)$. For simplicity, we call such samples $r$ suffering from high errors \emph{flawed samples}, and $L_p(r)$ \emph{flawed locations}.

\begin{problem}\label{prob1}
Given a localization model $\mathcal{L}$ learned from the training dataset $\mathbb{D}$, we want to optimize the localization errors of $\mathcal{L}$ on a testing dataset $\mathfrak{D}$, by ({1}) detecting those flawed samples $r\in\mathfrak{D}^- \sqsubseteq \mathfrak{D}$ and ({2}) repairing the flawed location $L_p(r)$.
\end{problem}

In the problem above, we say that a testing MR sample $r\in \mathfrak{D}^-$ is flawed and $L_p(r)$ is a flawed location if $||L_p(r)- L_t(r)||>\tau$ is met, where $\tau$ is a predefined threshold. We denote all flawed testing samples by $\mathfrak{D}^-$, and the normal testing MR samples by $\mathfrak{D}^+ = \mathfrak{D}-\mathfrak{D}^-$. In terms of the threshold $\tau$, it depends upon the localization error of $\mathcal{L}$ and used data set. For example, we tune $\tau$ by the 80\% error, 75 meters, of $\mathcal{L}$ in one of our used Jiading 2G data set. We will discuss the tuning of $\tau$ in Section \ref{s:evaluation}.

To solve the problem above, we have to tackle the following challenges. In the problem above, for one MR sample $r\in \mathfrak{D}$, if the true location $L_t(r)$ is available beforehand, we can comfortably determine whether or not the condition $||L_p(r)- L_t(r)||>\tau$ is met, and then find the flawed samples $\mathfrak{D}^-$. Yet, the testing MR samples $r\in \mathfrak{D}$ do not have the ground truth locations $L_t(r)$, and it is rather hard to determine or not the aforementioned condition is met and then to perform outlier detection and repair. Even if we can detect the flawed MR samples $r\in \mathfrak{D}^-$, how to repair flawed locations $L_p(r)$ is still non-trivial. Since the ground true location $L_t(r)$ is the most desirable one to repair $L_p(r)$, it is challenging to choose an appropriate location to replace $L_p(r)$ when the ground truth $L_t(r)$ is unavailable.

\subsection{Solution Overview}

\begin{figure}[h]
  \centering
  \includegraphics[width=3.6in]{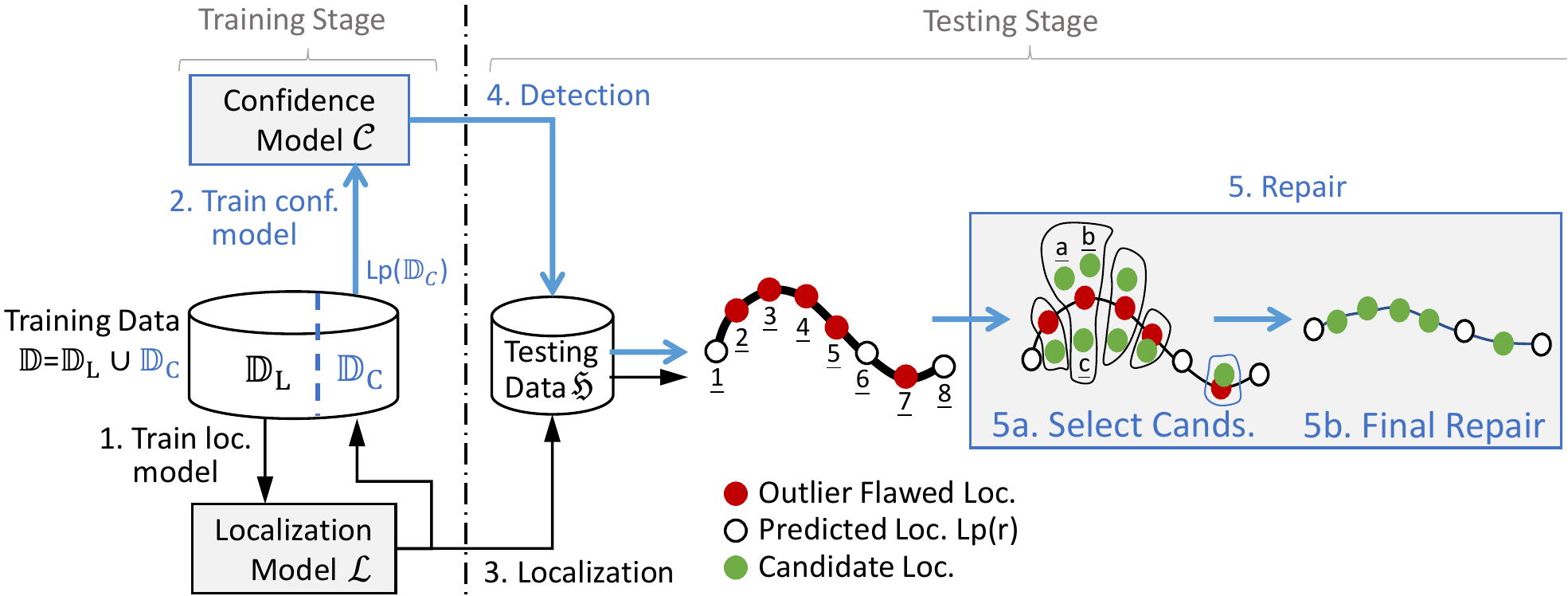}\vspace{-2ex}
  \caption{Overview of $\mathsf{RLoc}$} \label{fig:graph}
\end{figure}

To address the challenges above, the proposed solution \textsf{RLoc} essentially includes three components: a \emph{localization} model $\mathcal{L}$, an outlier \emph{detection} algorithm to find flawed samples $r\in\mathfrak{D}^-$ via a confidence model $\mathcal{C}$, and an outlier \emph{repair} algorithm to replace the flawed locations $L_p(r)$ by better ones. In terms of the localization model $\mathcal{L}$, we improve the previous work CCR \cite{ZhuLYZZGDRZ16} by using a classifier instead of the original regressor. The basic idea of the classifier is as follows. We first divide an area of interest to small square grid cells beforehand, and then build a classifier between MR samples (or equivalently MR features) and the grid cells where the GPS coordinates of MR samples are located. The classifier indicates a mapping function from MR features to grid cells. After the classifier is ready, we can predict one target grid cell which a testing MR sample belongs to, and take the centroid of the classified grid cell as the predicted location.

In Figure \ref{fig:graph}, \textsf{RLoc} involves training and testing stages. In the training stage, we divide the training data set $\mathbb{D}$ into two disjoint subsets: $\mathbb{D}_L$ and $\mathbb{D}_C$, i.e., $\mathbb{D} = \mathbb{D}_L\cup \mathbb{D}_C$. The first subset $\mathbb{D}_L$ is used to train the aforementioned localization model $\mathcal{L}$ (\emph{step 1}), and we apply $\mathcal{L}$ on the 
the second subset $\mathbb{D}_C$ to generate predicted locations $L_p(\mathbb{D}_C)$. With help of the prediction result $L_p(\mathbb{D}_C)$, we then train a sequence-based confidence model $\mathcal{C}$ (\emph{step 2}). In the testing stage, we again apply the localization model $\mathcal{L}$ on testing sequences of MR samples $r\in\mathfrak{D}$ to generate trajectories of predicted locations $L_p(r)$ (\emph{step 3}), and meanwhile apply the already trained confidence model $\mathcal{C}$ on the testing MR samples $\mathfrak{D}$ to detect flawed samples $\mathfrak{D}^-$ (\emph{step 4}). To correct the flawed positions of $\mathfrak{D}^-$, the repair algorithm first selects candidate locations, and then chooses the best ones to repair the flawed locations (\emph{step 5}).

Until now, we can find that \textsf{RLoc} significantly differs from traditional Telco localization. First, though we divide our approach into training and testing stages, the \emph{step 2} requires $L_p(\mathbb{D}_C)$, i.e., the prediction locations of the subset $\mathbb{D}_C$ by the model $\mathcal{L}$. We then exploit the prediction locations $L_p(\mathbb{D}_C)$ to acquire the labels of confidence levels, which are next used to train the confidence model $\mathcal{C}$ and finally to perform outlier detection and repair. Thus, we can intuitively treat the confidence-model-based outlier detection and repair (i.e., \emph{steps 2, 4, 5} in  Figure \ref{fig:graph}) as a \emph{post-processing} phase of traditional Telco localization. Second, in terms of the outlier detection and repair, the previous works such as CRL (Confidence model-based data Repairing technique for Telco Localization) \cite{ZhangRYZY17}, employ single-point-based detection and repair algorithms and do not take into account the connectivity of neighbouring locations. Instead, we adopt sequential detection and repair algorithms for better results. Finally, to guarantee the \emph{fairness} between our approach and other competitors, we still use $\mathbb{D}$ as the overall training dataset for the localization, detection and repair algorithms in \textsf{RLoc}, and $\mathfrak{D}$ as the testing dataset, with no extra training MR samples.

In the following Sections \ref{sec:hmm} and \ref{sec:rrepair}, we present the proposed detection and repair algorithms, respectively. Moreover, if without special mention, we by default say that the proposed detection/repair models are all sequence-based and MR samples have been re-processed to be sequence data.

%% file: 03-hmm.tex
\section{Confidence-based Detection Approach}\label{sec:hmm}
In this section, we first introduce the confidence level (Section \ref{sec:conf1}), and then present a sequence-based outlier detection algorithm via the proposed confidence model (Section \ref{sec:conf3}).

\subsection{Confidence Level}\label{sec:conf1}
We define the confidence level by a binary indicator. If the confidence level of a MR sample $r\in \mathfrak{D}$ is 0, the sample $r$ is flawed and otherwise normal.

\begin{definition}\label{def:conflevel}
For a MR sample $r$ and a localization model $\mathcal{L}$, if the distance $||L_p(r)-L_t(r)||$ between a prediction location $L_p(r)$ and ground truth $L_t(r)$ is greater than a predefined threshold $\tau$, i.e., $||L_p(r)-L_t(r)||> \tau$, then we say that $r$ is a flawed sample and the confidence level of $r$ is 0, and otherwise a normal sample with the confidence level 1.  
\end{definition}

To predict the confidence level of a testing sample $r$, our general idea is to learn a machine-learning-based confidence model that maps from training MR samples to the corresponding labels of confidence levels. Unfortunately, the original training dataset $\mathbb{D}$ only contains MR samples and GPS positions, but not confidence levels. To this end, we give the following steps to find the labels of confidence levels for training samples. Recall that we use the two disjoint subsets $\mathbb{D}_L$ and $\mathbb{D}_C$ to train a localization model $\mathcal{L}$ and a confidence model $\mathcal{C}$, respectively (see Figure \ref{fig:graph}). After the localization model $\mathcal{L}$ is trained by $\mathbb{D}_L$, we then apply $\mathcal{L}$ on the subset $\mathbb{D}_C$ to predict the locations $L_p(r)$ for the sample $r\in \mathbb{D}_C$. Since $\mathbb{D}_C$ is still a training data subset, the sample $r\in \mathbb{D}_C$ has the ground truth position $L_t(r)$. We then follow Definition \ref{def:conflevel} to compute the confidence level for every sample $r\in \mathbb{D}_C$. Once the confidence level is available, we train a machine-learning-based confidence model $\mathcal{C}$ from these samples $r\in \mathbb{D}_C$ to corresponding confidence levels. After that, we apply the trained model $\mathcal{C}$ on testing samples $\mathfrak{D}$ to detect flawed ones $\mathfrak{D}^-$.

In terms of the specific machine learning algorithm used to train the confidence model $\mathcal{C}$, a simple approach is to exploit a binary-classifier such as Random Forest or GBDT (Gradient Boosting Decision Tree) \cite{gdbt} to learn the mapping function from an individual sample $r\in \mathbb{D}_C$ to its confidence level. Note that it is straightforward to extend our binary confidence levels to a multi-level confidence model (e.g., using the levels from 1 to 5). For example, we could leverage a multi-classifier, instead of a binary classifier, to support the multi-level confidence model. 

Nevertheless, the approach above does not take into account the underlying spatio-temporal locality in neighbouring MR samples, and is still inaccurate. In the rest of this section, to capture the underlying spatio-temporal locality in neighbouring samples, we estimate the confidence levels of MR sequences for higher accuracy first via a static HMM confidence model and then via an improved one, namely DA-HMM.

\subsection{Static HMM-based Confidence Model}\label{sec:conf3}
In this section, we train a static HMM-based confidence model $\mathcal{C}$ to learn the mapping between each MR sequence in $\mathbb{D}_C$ and a sequence of confidence levels by the following intuition.

Let us consider the scenario: a mobile device is moving first close to a certain serving base station (say $bs$) and then far away from $bs$, until the device is with another serving base station. In this scenario, the mobile device generates a sequence of MR samples. The signal strength of $bs$ within such MR samples becomes first stronger and later weaker. If we treat the signal strength $ss$ (e.g., RSSI) of $bs$ in MR samples as \emph{observation} and the confidence level as \emph{state}, then the states (i.e., confidence levels) first become greater (i.e., one) and next smaller (i.e., zero).

When given an observed sequence of MR samples (containing $bs$ and $ss$), we expect to infer a corresponding sequence of confidence levels via the following HMM decoding problem: given the parameters of HMM (acquired from the training data $\mathbb{D}_C$) and the MR observation sequence for the testing dataset $\mathfrak{D}$, we aim to find the most likely sequence of states (confidence levels). Formally, we describe the static HMM $\lambda =(S, V, A, B, \pi)$ as follows.

\begin{itemize}
 \item $S =\left\{0, 1 \right\}$ is the set of states (confidence levels).
\item $V = \left\{v_1,...,v_k,...,v_M \right\}$ is the set of observations $v_k = \langle v_k^{bs}, v_k^{ss}\rangle$, where $v_k^{bs}$ is a list of up to 7 base stations $bs$ and $v_k^{ss}$ is the list of associated $ss$. Moreover, we convert the continuous readings of $ss$ into 8 discrete levels: $ss$ within the range $[-50, -110]$ is converted to 6 levels from 2, 3,..., to 7 by the equal interval of length 10, $ss<-50$ and $ss>-110$ to the levels of 1 and 8, respectively. 
%Note that the $v_k^{bs}$ of 4G MR data collected by frontend Android devices is a list of only one serving base station, but its $v_k^{ss}$ contains up to 7 signal levels of $ss$, e.g., $v_k^{bs}=\left\{BS_{A}\right\}$ and $v_k^{ss}=\left\{2,3,4,4,4,6,7\right\}$ (the details of this dataset will be discussed in Section \ref{sec:experimental setting}).
\item $A = \left\{a_{ij}\right\}$ is the distribution of state transition probability $a_{ij}$ of going from the confidence level $i$ at time step $t$ to the next confidence level $j$ at time step $t+1$.
\item $B = \left\{b_j(k)\right\}$ is the distribution of emission probability $b_j(k)$ of observation $v_k$ in state $j$.
 \item $\pi= {\pi_i}$ is the initial state distribution with $\pi_i =P[q_1 = S_i]$.
\end{itemize}

\if 0
\begin{table}[h]
\scriptsize
\begin{tabular}{|c|c|c|c|c|c|c|c|c|}
\hline
RSSI (dbm)   & $\geq$-50 & (-50,-60{]} & (-60,-70{]} & ... & (-100,-111{]} & no signals \\ \hline
\hline
$v_k^{ss}$ & 1 & 2  & 3  & ... & 7  & 8  \\\hline
\end{tabular}
\caption{RSSI Reading Levels.}  \label{tab:rssi}
\end{table}

In the offline training phase, we train the static HMM model $\lambda =(S, V, A, B, \pi)$ based on the statistics of the subset $\mathbb{D}_C$ and associated confidence levels. In the online detection phase, with help of the trained HMM model, we adopt the Viterbi algorithm \cite{Viterbi1967Error} to predict the most likely sequence of confidence levels, i.e., states $Q = \left\{q_1,...,q_{t} \right\}$, from a sequence of observations $O = \left\{o_1,...,o_{t} \right\}$ in an input sequence of testing MR samples $\mathfrak{D}$.
\fi

\subsection{A Dual-Stage Adaptive HMM}\label{ssec:da-hmm}
The static HMM model above may not work well on real MR samples: the neighbouring MR samples within real sequence data frequently exhibit uncertain timestamp intervals, e.g., caused by various sampling rate and data missing. Thus, besides the states $s_i$ and $s_j$, the state transition probability $a_{ij}$ further depends upon the timestamp intervals between neighbouring MR samples. Moreover, due to the high cost of collecting training samples, it is not rare that some areas of interest suffer from insufficient samples, leading to inaccurate estimation of the emission probability $b_j(k)$.

To address the issues above, we propose a dual-stage adaptive HMM, named DA-HMM, on top of the static HMM model. Specifically, after a static HMM model is learned by the samples $\mathbb{D}_C$, in the training phase of DA-HMM, we introduce the time interval $\Delta$ between neighbouring MR samples and the sample size $\gamma$ for observation $k$ in state $j$, and define the new transition probability $a_{ij}^\Delta$ and emission probability $b_j^{\gamma}(k)$, respectively. The new probabilities are then adaptive to $\Delta$ and $\gamma$. The detail to estimate  $a_{ij}^\Delta$ and $b_j^{\gamma}(k)$ is as follows.

\subsubsection{Adaptive State Transition Probability}\label{sec:astate}

\begin{figure}[th]
%	\hspace{-10ex}
	\begin{center}
		\begin{tabular}{c c c}
			\begin{minipage}[t]{0.3\linewidth}
				\begin{center}
					\centerline{\includegraphics[totalheight=0.85in]{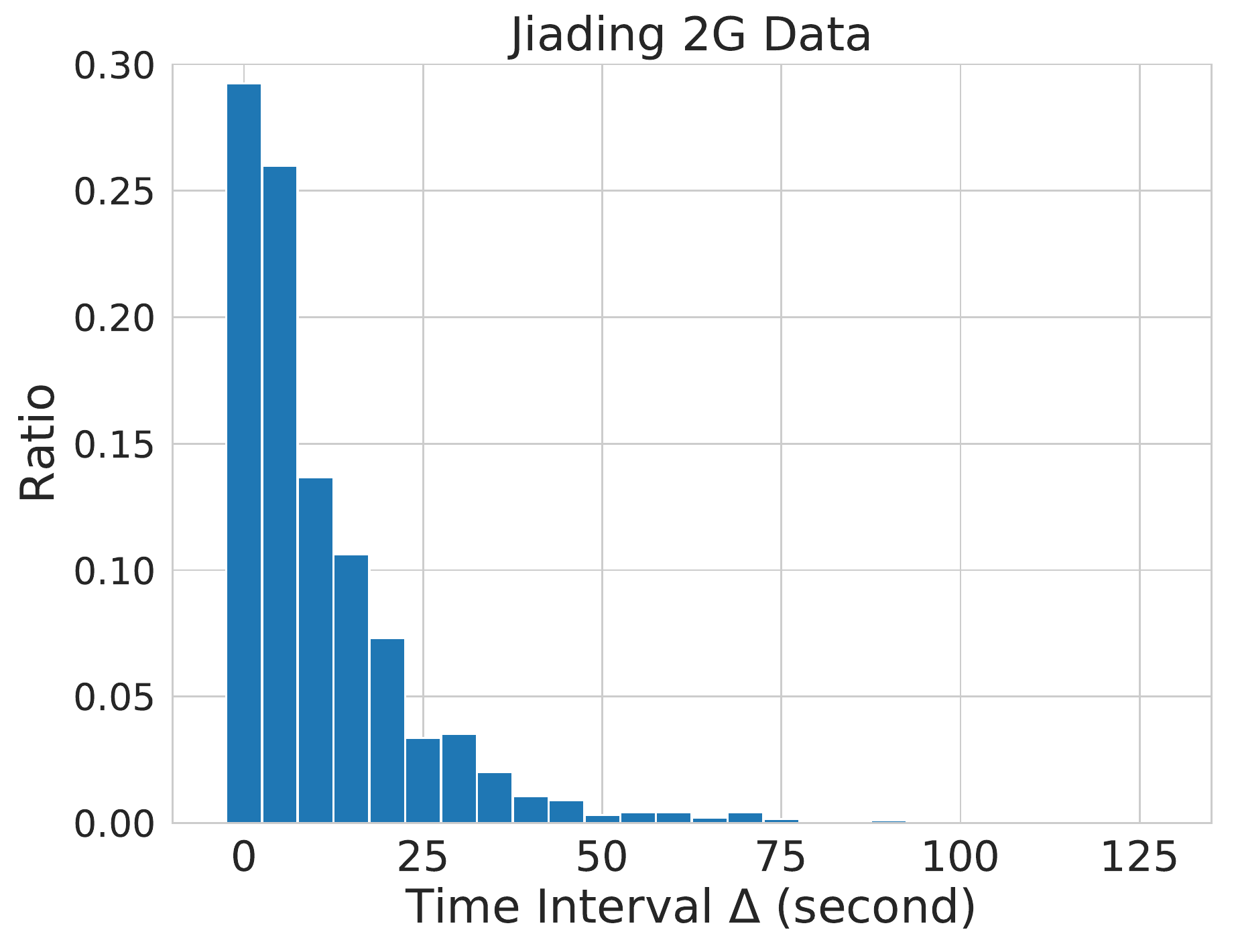}}
				\end{center}
%                \caption{Time Interval Distribution. (Jiading 2G Data Set)} \label{fig:time_dis}
			\end{minipage}
			&
			\begin{minipage}[t]{0.3\linewidth}
				\begin{center}
					\centerline{\includegraphics[totalheight=0.85in]{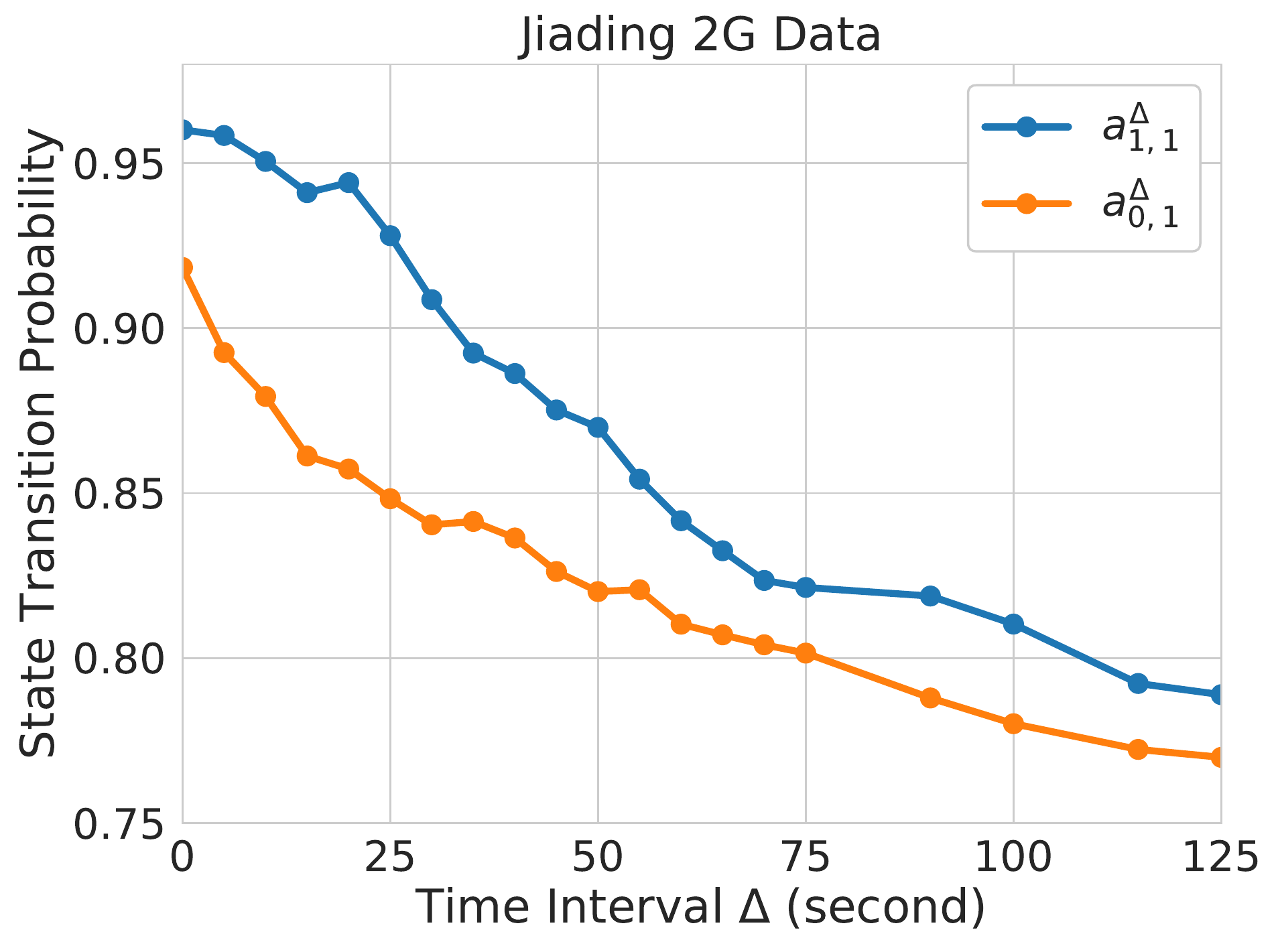}}
				\end{center}
                % \caption{State Transition Probability $P_{s_i=1 \rightarrow s_j=1}^{\Delta T}$ under Different Time %Intervals.}  \label{fig:prob_change}
			\end{minipage}
            &
			\begin{minipage}[t]{0.3\linewidth}
				\begin{center}
					\centerline{\includegraphics[totalheight=0.85in]{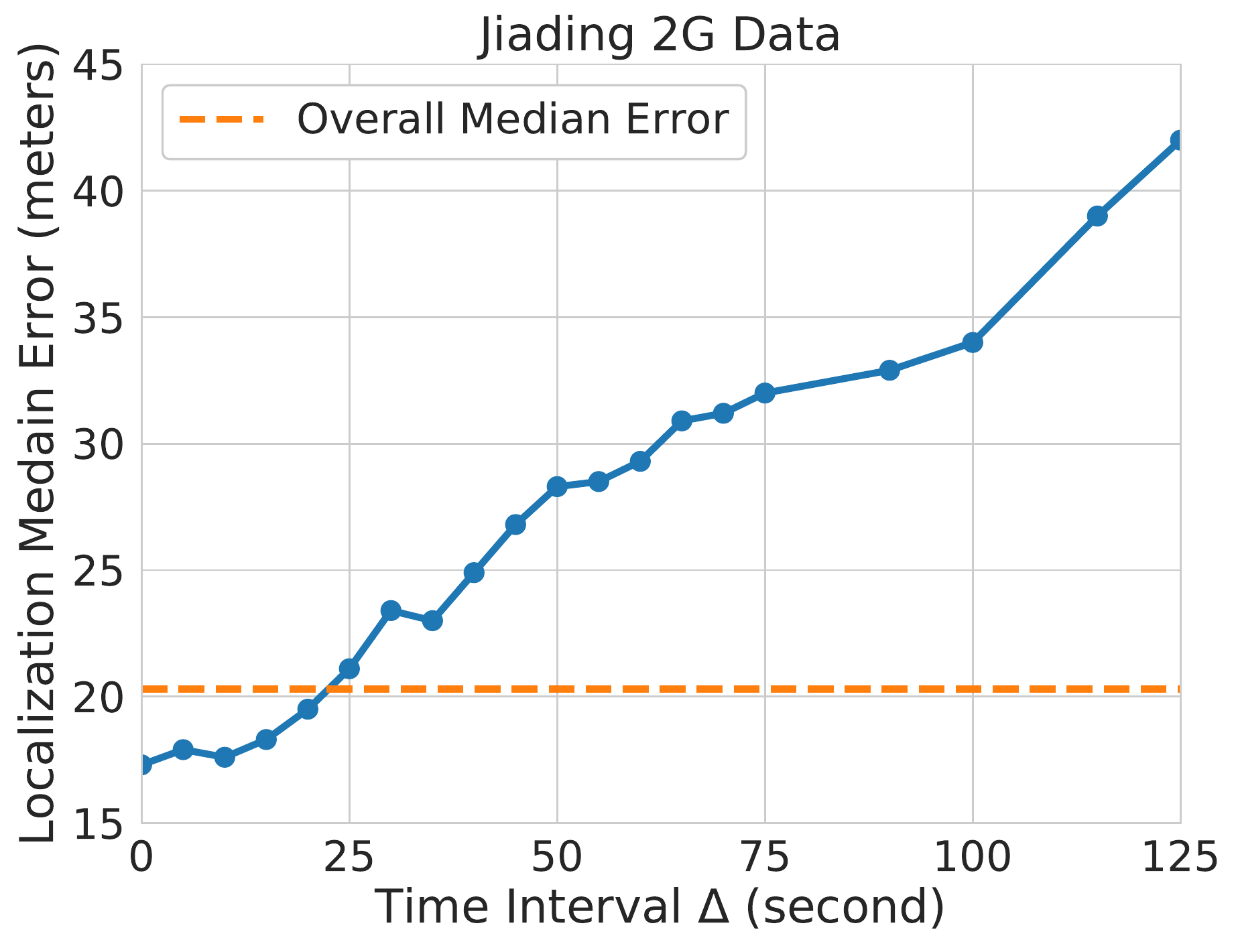}}
				\end{center}
                % \caption{Variation Trend of Localization Median Error under Different Intervals.}  %\label{fig:prob_delta}
			\end{minipage}
            %&
			%\begin{minipage}[t]{0.25\linewidth}
				%\begin{center}
					%\centerline{\includegraphics[height=1.45in]{fig/delta_state_transition.eps}}
				%\end{center}
                 %\caption{Deviation $|P_{s_i=1 \rightarrow s_j=1}^{\Delta T}-\overline{P}_{s_i=1 \rightarrow s_j=1}|$ under Different Time Intervals. (Jiading 2G Data Set)}  \label{fig:prob_delta}
			%\end{minipage}
		\end{tabular}\vspace{-3ex}
	\end{center}
                 \caption{From left to right: (a) Time Interval Distribution, (b) State Transition Probability $a_{i,1}^{\Delta}$, (c) Localization Median Error.}  \label{fig:time_dis}
\end{figure}

Figure \ref{fig:time_dis} first motivates the design of the adaptive state transition probability. The leftmost figure indicates a rather uneven distribution of timestamp intervals among neighbouring MR samples: the intervals vary from 0 and 125 seconds, instead of a fixed value. In the middle figure (we omit the curves of $a_{1,0}$ and $a_{0,0}$ due to $a_{1,0} = 1.0 - a_{1,1}$ and $a_{0,0} = 1.0 - a_{0,1}$), the probabilities $a_{0,1}$ and $a_{1,1}$ become decreased with a greater timestamp interval $\Delta$. The rightmost figure gives the localization error under various timestamp intervals. Greater timestamp intervals indicate higher localization errors and vice versa. It makes sense: a smaller timestamp interval means densely sampled MR data, leading to higher spatio-temporal locality and thus smaller localization errors. As a summary, Figure \ref{fig:time_dis} clearly indicates that the probabilities $a_{1,1}$ and $a_{0,1}$ significantly vary from  timestamp intervals $\Delta$ and thus using a fixed interval does not work well.

To design the adaptive state transition probability $a_{i,1}^{\Delta}$  (where $i$ = 1 or 0), we first note that $a_{i,1}^{\Delta}$ decreases by a greater timestamp interval $\Delta$ (see Figure \ref{fig:time_dis}b). To this end, we exploit an exponential decrease-based time decay model which has been widely used for mobility pattern analytic and usually treated as an exponential regression mode \cite{DBLP:journals/tmc/KaragiannisBV10,DBLP:journals/amc/CV19}.
\begin{equation}\scriptsize
\begin{aligned}
a_{i,1}^{\Delta} &= e^{(-\alpha_i\Delta)} \cdot \beta_{i}\cdot{a}_{i,1}%\quad;\quad\quad a_{i,0}^{\Delta} = 1-a_{i,1}^{\Delta}
\end{aligned}
\label{eq:state_transition}
\end{equation}

To derive the $a_{i,1}^{\Delta}$ above, we need to estimate $\alpha_i$ and $\beta_i$ from the training data subset $\mathbb{D}_C$. That is, for each discrete time interval $\Delta$ in $\mathbb{D}_C$, we estimate $a_{i,1}^{\Delta,true}$ via the statistics of $\mathbb{D}_C$.
\begin{equation}\scriptsize
\begin{aligned}
a_{i,1}^{\Delta,true} = \frac{U_{i,1}^{\Delta}}{U_{i,1}^{\Delta}+U_{i,0}^{\Delta}}\\
\end{aligned}
\label{eq:trans_pro_estimate}
\end{equation}
where $U^{\Delta}_{i,j}$ denotes the count of the training samples in $\mathbb{D}_C$ that satisfy \emph{1)} the hidden state of the sample is $i$ at time step $t$, \emph{2)} the hidden state transfers to $j$ at time step $t+1$, and \emph{3)} the time interval between time steps $t$ and $t+1$ is $\Delta$. Next, we exploit the Gauss-Newton algorithm \cite{myers1990classical} to finally estimate $\alpha_i$ and $\beta_{i}$ with help of $\Delta$ and $a_{i,1}^{\Delta,true}$.

\subsubsection{Adaptive Emission Probability}\label{sec:aemission}
Given the observation $v_k = \langle v_k^{bs}, v_k^{ss}\rangle$, we first estimate the static emission probability $b_{j}(k)$ by the statistics of  the training data subset $\mathbb{D}_C$ as follows.

\begin{equation}\scriptsize
\begin{aligned}
{b}_j(k)={P}(v_{k}^{bs},v_{k}^{ss}|s_j) &= \frac{P((v_k^{bs},b_k^{ss}),s_j)}{P(s_j)} \\
      &\approx	\frac{|\mathbb{D}_C(v_{k}^{bs})  \cap  \mathbb{D}_C(v_{k}^{ss})  \cap    \mathbb{D}_C(s_j)|}{|\mathbb{D}_C(s_j)|}
\end{aligned}
\label{eq:em_pro}
\end{equation}

In the equation above, we estimate the probability $P(s_j)$ by using the carnality of set $\mathbb{D}_C(s_j)$, i.e., the count of MR samples within $\mathbb{D}_C$ involving state $s_j$. Since the state (i.e., confidence level) is represented by a binary indicator, we reasonably assume that $\mathbb{D}_C(s_j)$ could contain sufficient samples, then the estimation of ${P}(s_j)$ above makes sense.

Yet, to estimate $P((v_k^{bs},b_k^{ss}),s_j)$ in the numerator, we have to find the carnality $|\mathbb{D}_C(v_{k}^{bs}) \cap \mathbb{D}_C(v_{k}^{ss})  \cap  \mathbb{D}_C(s_j)|$. Unlike $v_k^{ss}$ and $s_j$, the base stations in $v_k^{bs}$ are uniquely identified and $v_{k}^{bs}$ may be a list of base stations that are rarely sampled within $\mathbb{D}_C$. Thus, the estimation of $P((v_k^{bs},b_k^{ss}),s_j)$ is rather sensitive to $|\mathbb{D}_C(v_{k}^{bs})|$. In case that the carnality $|\mathbb{D}_C(v_{k}^{bs})|$ is smaller than the aforementioned threshold $\gamma$, Equation \ref{eq:em_pro} may not precisely estimate $P((v_k^{bs},b_k^{ss}),s_j)$.

To overcome the issue above, our basic idea is to leverage those MR observations $v_{k'}$ such that $v_{k}$ and $v_{k'}$ are similar in terms of the Jaccard similarity coefficient between $v_{k'}^{bs}$ and $v_{k}^{bs}$, i.e., the similarity $J(v_{k}^{bs}, v_{k'}^{bs})\geq \varepsilon$, where $\varepsilon$ is a given threshold $\varepsilon$.
Then, for every similar observation $v_{k'}$, we define a weight $w_{k'}$ and give a weighted adaptive emission probability

\begin{equation}\scriptsize\begin{aligned}
b^{\gamma}_{k}(k) =
\frac{{P}^{\gamma}((v_{k}^{bs},v_{k}^{ss}),s_j)}{P(s_j)} \approx&
 \sum_{k'} w_{k'}\cdot \frac{|\mathbb{D}_C(v_{k'}^{bs})  \cap  \mathbb{D}_C(v_{k}^{ss}  \cap    \mathbb{D}_C(s_j)|}{|\mathbb{D}_C(s_j)|} \\
 \mbox{where}\ w_{k'}=&\frac{\lg (1+z_{v_{k'}^{bs}})\cdot J(v_{k}^{bs}, v_{k'}^{bs})}
{\sum_{v_{d}}{\left(\lg (1+z_{v_{d}^{bs}})\cdot J(v_{k}^{bs}, v_{d}^{bs})\right)}}
\end{aligned}
\label{eq:em_pro_new}
\end{equation}

In the equation above, $z_{v_k}^{bs} = |\mathbb{D}_C(v_{k}^{bs})|$ denotes the size of those training MR samples $\mathbb{D}_C(v_{k}^{bs})$ and $v_d$ denotes every observation similar to $v_k^{bs}$. Note that such similar observations represented by $v_d$ are actually those represented by $v_{k'}$ and we introduce the notation $v_d$ just to avoid confusion between $v_d$ and $v_{k'}$ in the equation above. In this way, by introducing the sum in the denominator, $w_{k'}$ is a normalized weight. Here, the item $\log{(1+z_{v_{k'}^{bs}})}$ ensures a valid logarithmic operation even for $z_{v_{k'}^{bs}}=0$. The intuition of weight $w_{k'}$ is as follows. When more training samples ${\mathbb{D}_C}(v_{k'}^{bs})$ (a.k.a a greater size $z_{v_{k'}^{bs}}$) have the observation $v_{k'}$ (which is similar to $v_{k}$ with the coefficient $J(v_{k}^{bs}, v_{k'}^{bs}) \geq \varepsilon$), we have a greater weight $w_{k'}$. 
%Note that the $v_k^{bs}$ of 4G MR samples collected by Android devices only contains one serving base station, thus there is no need to find its similar base station observation $v_{k'}^{bs}$ and the calculation of associated adaptive emission probability follows Equation \ref{eq:em_pro}.

\begin{figure}[ht]
	\centering
	\includegraphics[width=2.0in]{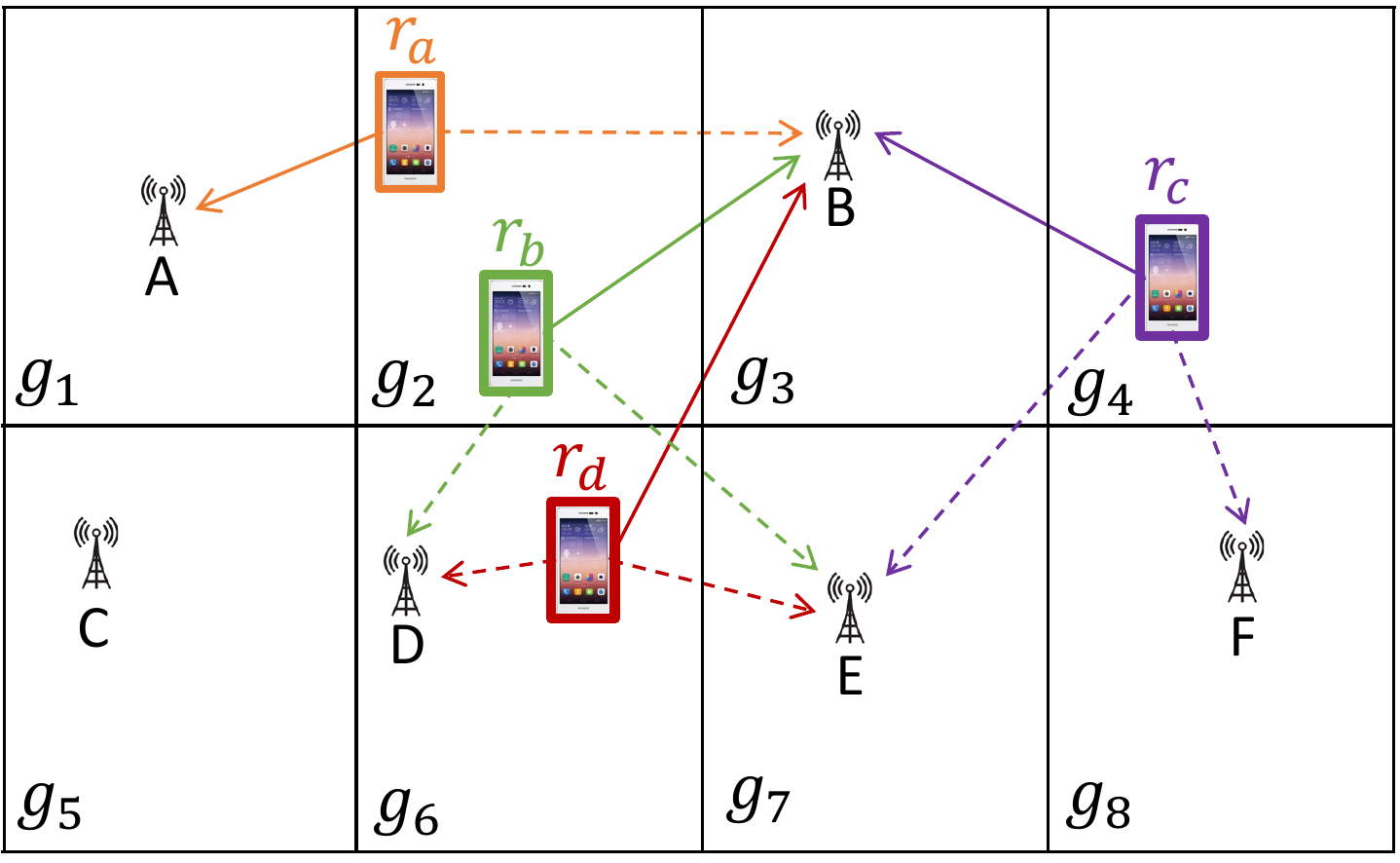}\vspace{-2ex}
	\caption{Example of 8 grid cells $g_1...g_8$, 6 base stations $A...F$ and 4 MR samples $r_a...r_d$. Solid (resp. dotted) lines indicate the connection of mobile devices to serving (resp. non-serving) stations.} \label{fig:8grids}
\end{figure}

\begin{example}
For simplicity, we assume that the four training MR samples in Figure \ref{fig:8grids} are all with the exactly same Telco signal strength observation $v_k^{ss}$ (e.g., all RSSI levels of these samples are 2) and the states of these samples are all 0 except that the state of $r_c$ is 1. Thus, we directly remove the subitems with respect to $v_k^{ss}$ in the estimation of the emission probability (see Equations \ref{eq:em_pro} and \ref{eq:em_pro_new}).

Suppose that we have the thresholds $\gamma = 2$ and $\varepsilon = 0.5$ and need to estimate the emission probability $b_j(k)$ for $s_j=0$ and $v_k^{bs}=\{B,E,F\}$. Since no sample is with such $s_j$ and $v_k^{bs}$, we could follow Equation \ref{eq:em_pro} and estimate the static emission probability by zero. Nevertheless, this estimation, which is sensitive to the sample size of $\mathbb{D}_C(v_k^{bs})$ with $|\mathbb{D}_C(v_k^{bs})|<\gamma = 2$, may not make sense.

Instead we follow Equation \ref{eq:em_pro_new} to find two similar observations $v_{k'}^{bs}$: $\{B,D,E\}$ in samples $r_b$ and $r_d$, and $\{B,E,F\}$ in sample $r_c$. Then for the first observation $v_{k'}^{bs} = \{B,D,E\}$, we can compute the Jaccard similarity $J(v_{k}^{bs},v_{k'}^{bs})=0.5$ and next $w_{k'}=\frac{\lg(2+1)\cdot0.5}{\lg(2+1)\cdot 0.5+\lg(1+1)\cdot1}=0.44$. For the second observation $v_{k'}^{bs} = \{B,E,F\}$, we compute $w_{k'}=0.56$. Finally, we estimate $b^{\gamma}_j(k)=0.44\cdot \frac{2}{3}+0.56\cdot \frac{0}{3}$ = 0.293.

Recall that among the base stations within 4G LTE MR samples collected by Android devices, only the serving station is valid and other stations might be null (see Section \ref{sec:bkmrdata}). Then, to estimate the emission probability $b_j(k)$ for $s_j=0$ and $v_k^{bs}=\{B\}$ (differing from the above $v_k^{bs}=\{B,E,F\}$), we have $|\mathbb{D}_C(v_k^{bs})|=3>\gamma = 2$ and then follow Equation \ref{eq:em_pro} to compute $b^{\gamma}_j(k) \approx	\frac{|\mathbb{D}_C(v_{k}^{bs})  \cap  \mathbb{D}_C(v_{k}^{ss}) \cap   \mathbb{D}_C(s_j)|}{|\mathbb{D}_C(s_j)|}=\frac{2}{3}=0.667$.

%Secondly, since $|\mathbb{D}_C|$ equals to 4 with %$s_j=0$, we estimate %$P((v_k^{bs},v_k^{ss}),s_j)=\frac{0}{4}=0$ and %$P((v_{k'}^{bs},v_k^{ss}),s_j)=\frac{2}{4}=0.5$, %respectively.

%Thirdly, we estimate %$P^{\gamma}((v_k^{bs},v_k^{ss}),s_j)=0.22$ by the sum of %weighted probabilities $w_k\cdot %P((v_k^{bs},v_k^{ss}),s_j)=0.56\cdot0=0$ and $w_{k'}\cdot %P((v_{k'}^{bs},v_k^{ss}),s_j)=0.44\cdot0.5=0.22$.
% Finally, adaptive emission probability is estimated as %$b^{\gamma}_j(k)=\frac{P^{\gamma}((v_k^{bs},v_k^{ss}),s_j%)}{P(s_j)}=\frac{0.22}{0.75}=0.293$.
%\end{enumerate}
\end{example}

To summarize the steps above, in Algorithm \ref{al2g:DA-HMM}, we give the Pseudo-code to estimate the parameters of DA-HMM. First, the lines 1-4 follow Section \ref{sec:astate} to estimate the adaptive state transition probability $a_{i,j}^{\Delta}$, and lines 6-22 follow
Section \ref{sec:aemission} to estimate the adaptive emission probability $b_{j}^{\gamma}(k)$.

\begin{algorithm}\scriptsize
    \caption{Parameter Estimation in DA-HMM}\label{al2g:DA-HMM}
    \KwIn{Static HMM model $\lambda =\{{S},{V},{A},{B},{\pi}\}$, Training MR subset $\mathbb{D}_C$, Thresholds $\gamma$ and $\varepsilon$}
    \KwOut{$a_{i,j}^{\Delta}$ and $b_{j}^{\gamma}(k)$}
    Create a time interval list $TList$ from nbr. samples within MR sequences in $\mathbb{D}_C$\;
    \lForEach {time interval $\delta \in TList$}
    {
        Infer $a_{i,1}^{\delta,true}$ by the statistics of $\mathbb{D}_C$
    }
    Estimate the parameters $\alpha_a$ and $\beta_i$ in Eq. (\ref{eq:state_transition}) with $\delta$ and $a_{i,1}^{\delta,true}$ by \emph{Gaussian-Newton} method\;
    Update $a_{i,1}^{\Delta} \leftarrow e^{(-\alpha_i\Delta)} \cdot \beta_{i} \cdot a_{i,1}$ and $a_{i,0}^{\Delta} \leftarrow 1-a_{i,1}^{\Delta}$ \;

    \ForEach{observation $v_{k}$ in $V$}
    {
        Compute the sample size $z_{v_k^{bs}} \leftarrow |\mathbb{D}_C(v_{k}^{bs})|$\;
        \lIf{$z_{v_k^{bs}} \ge \gamma$}{ 
        Compute $b^{\gamma}_j(k)$ by Eq. (\ref{eq:em_pro})}
        \Else{
            \ForEach{similar observation $v_{k'}^{bs}$ with $J(v_{k}^{bs}, v_{k'}^{bs})\ge \varepsilon$}
            {
                $z_{k'}^{bs} \leftarrow |\mathbb{D}_c(v_{k'}^{bs})|$, and Compute $w_{k'}$ by Eq. (\ref{eq:em_pro_new})\;
                
            }
             Compute$b^{\gamma}_j(k)$ by Eq. (\ref{eq:em_pro_new})\;

        }
        
    }
    \Return $a_{i,j}^{\Delta}$ and $b_{j}^{\gamma}(k)$ \;
\end{algorithm}

%% file: 04-repair.tex
\section{Location Repair}\label{sec:rrepair}
Recall that the proposed confidence model can be applied onto a testing sequence $\mathcal{R}=\{r_1,...,r_{|R|}\}\sqsubseteq \mathfrak{D}$ of MR samples to detect flawed samples. We are interested whether or not these flawed samples are neighboring within the sequence $\mathcal{R}$. For example, in Figure \ref{fig:graph}, we have detected five flawed samples within an input sequence of 8 testing samples. Four of them (i.e., $2,...,5$) are neighbouring within the input sequence and yet the one $7$ is disjoint from all other flawed samples. For a disjoint flawed sample, we find the most appropriate candidate location to replace the flawed location $L_p(r)$. Instead, to repair the neighboring flawed locations, we then find the best sequence (a.k.a trajectory) of candidate locations. Since an individual flawed location can be treated as the special case of a sequence with the sequence length equal to 1, we thus generally focus on \emph{1}) finding the candidate locations for every flawed location and \emph{2}) the repair of an entire sequence of neighboring flawed locations.

\subsection{Candidate Positions}\label{sec:candpos}
Recall that our multi-classifier-based localization model has already divided an area of interest into multiple small grid cells. Thus, to find candidate positions, we alternatively select candidate grid cells. Before giving the detail, we first give the following notations. For a certain flawed sample $r$, the notation $bs_r$ indicates the set of those base stations appearing in $r$, and $g_r$ denotes the grid cell where the position $L_p(r)$ is located. For a grid cell $g$, the notation $BS_g$ means the set of all base stations appearing in entire MR samples located within $g$.

With help of the notations above, we give the intuition of finding candidate positions. For a flawed sample $r$, a certain grid cell $g$ becomes the candidate of $g_r$, if the similarity of the two sets $bs_r$ and $BS_g$ is high and greater than a predefined threshold $\xi$. We measure the similarity as follows. Recall that $bs_r$ contains up to 6 or 7 base stations, i.e., $|bs_r| = 6 \sim 7$. Next, the grid cell $g$ may contain many MR samples and $|BS_g|$ is thus possibly much greater than $|bs_r|$. The standard Jaccard coefficient between $bs_r$ and $BS_g$, which is very close to 0.0 no matter $bs_r$, does not work well. Thus, we define a variant coefficient $J'(bs_{r}, BS_{g})=\frac{|bs_{r}\bigcap BS_{g}|}{|bs_{r}|}$.
%In addition, we expect that the candidate grid $g$ contains at least one MR sample $r'$ such that $bs_{r'}=bs_r$. It means that $r$ and $r'$ share the same base stations but possibly with very different Telco signal strength.
Based on the intuition above, we then give the following rule to find candidate grid cells for $g_r$.
\begin{equation}
\scriptsize
\begin{aligned}
%G_r=\{g\in \mathbb{G}|J'(bs_{r}, BS_{g})\geq \xi \  \mathbf{and}\   \exists	r'\in R_g \ \mathbf{with} \ bs_{r} = bs_{r'}  \}
G_r=\{g \in \mathbb{G}|J'(bs_{r}, BS_{g})\geq \xi \}
\end{aligned}
\label{eq:candidate_select}
\end{equation}
where $\mathbb{G}$ denotes the set of all spatial grid cells in the area of interest, and $\xi$ is a predefined threshold.

\begin{example}
%We still use Figure \ref{fig:8grids} for illustration where $r_d$ is a flawed sample and the threshold $\xi = 0.5$. For the grid cell $g_2$ with the set $BS_{g_2}=\{A, B, D, E\}$, we have $J'(bs_{r_d}, BS_{g_2}) = 1.0>\xi$. Meanwhile the grid $g_2$ contains a sample $r_b$ with $bs_{r_b} = bs_{r_d}$. We thus choose $g_2$ as a candidate. However, for the grid cell $g_r$, though with $J'(bs_{r_d}, BS_{g_4}) = 2/3>\xi$, we cannot find a sample $r'\in R(g_4)$ with $bs_{r'}=bs_{r_d}$. Thus $g_4$ cannot become a candidate.
Still in Figure \ref{fig:8grids}, we assume that $r_d$ is a flawed sample and the threshold $\xi = 0.5$. For the grid cell $g_2$ with the set $BS_{g_2}=\{A, B, D, E\}$ and $g_4$ with the set $BS_{g_4}=\{B, E, F\}$, we have $J'(bs_{r_d}, BS_{g_2}) = 1.0$ and $J'(bs_{r_d}, BS_{g_4}) = 2/3$, both of which are greater than $\xi$. We thus choose $g_2$ and $g_4$ as two candidates.
\end{example}

\subsection{Sequence-based Repair}\label{sec:repair1}
When given a sequence of flawed locations $L_p(r)$, the proposed repair algorithm considers 1) the \emph{possibility} or weight of a candidate grid to repair every flawed location and 2) the \emph{transition} possibility between two candidate grids, i.e., the possibility of mobile devices to move from one candidate grid to the next one. To this end, we propose to maximize the joint probability of the path to connect a sequence of candidate grids that are used to repair the entire sequence of flawed locations. Before giving the definition of the joint probability, we first define a \emph{repair graph}.

\emph{Repair Graph}: Consider a sequence $\mathcal{R}$ of $N (=|\mathcal{R}|)$ neighbouring flawed locations $L_p(r_i)$ with $1\leq i\leq N$. For each flawed location $L_p(r_i)$ and corresponding grid cell $g_{r_i}$, we have a set $G_i$ of at most $k$ candidate grids $g_{i,j}\in G_i$ with $1\leq j\leq k$. Formally, we define a \emph{repair graph} $\mathcal{G}$, where each vertex in $\mathcal{G}$ is mapped to a candidate grid $g_{i,j}$. We build a directed edge from a candidate vertex $g_{i,j}$ to another vertex $g_{i+1,j'}$, if the corresponding locations $L_p(r_i)$ and $L_p(r_{i+1})$ are neighbouring within the sequence $\mathcal{R}$. Each vertex (and edge) is with an associated weight or probability (we will give the probability soon). Given the graph $\mathcal{G}$, we have at most $k^N$ paths from the source to sink. Among all such paths, we want to find one path which is with the maximal joint probability to repair the $N$ flawed locations.

\begin{definition}[Vertex Weight]
For a flawed location $L_p(r_i)$, we define the vertex weight $W_{g_{i,j}}$ of a candidate grid $g_{i,j}$ to measure the goodness of $g_{i,j}$ to repair $L_p(r_i)$.
\end{definition}

For a non-flawed sample $r_i$, no candidate grid is needed and we simply set the vertex weight of $g_{r_i}$ by 1.0. For a flawed location $L_p(r_i)$ and a candidate $g_{i,j}$, we compute the vertex weight $W_{g_{i,j}}$ by the following equation.

\begin{equation*}
\scriptsize
\begin{aligned}
W_{g_{i,j}} = \frac{J'(bs_{r_i}, BS_{g_{i,j}}) \cdot P(g_{i,j}|BS_{g_{i,j}}) \cdot \exp({-D(bs_{r_i}^{1},BS_{g_{i,j}}^1)})}{\sum_{g \in G_{r_i}} J'(bs_{r_i}, BS_{g}) \cdot P(g|BS_{g}) \cdot \exp(-D{(bs_{r_i}^{1},BS_{g}^1)})}
\end{aligned}
\label{eq:candidate_weight}
\end{equation*}

In the equation above, we compute three sub-items.
\begin{itemize}
\item $J'(bs_{r_i},BS_{g_{i,j}})$: the similarity coefficient between the flawed MR ${r_i}$ and candidate grid cell $g_{i,j}$ in terms of their base stations.
\item  ${P}(g_{i,j}|BS_{g_{i,j}})$: the posterior probability.
\item $D(bs_{r_i}^{1},BS_{g_{i,j}}^1)$: the average physical distance between the serving base station $bs_{r_i}$ in MR sample $r_i$ and those serving stations of MR samples within $g_{i,j}$. Since the serving base station plays a key role in Telco localization, we thus introduce $D(\cdot)$ to compute $W_{g_{i,j}}$.

\begin{equation*}\scriptsize
\begin{aligned}
\scriptsize
D(bs_{r_i}^{1},BS_{g_{i,j}}^1)=\left\{
\begin{array}{lr}
0, &bs_{r_i}^1 \in BS_{g_{i,j}}^1\\
\frac{\sum_{bs \in BS_{g_{i,j}}^1}d(bs_{r,i}^1,bs)}{|BS_{g_{i,j}}^1|}, &\mbox{otherwise}\\
\end{array}
\right.
\end{aligned}\label{eq:bs_distance}
\end{equation*}
where $d(\cdot)$ denotes the Euclidean distance between the two base stations. Thus, $D(bs_{r_i}^{1},BS_{g_{i,j}}^1)$ indicates the average distance between the serving base station $bs_{r_i}$ and each serving station within the grid $g_{i,j}$.

\end{itemize}

Besides the vertex weight in a repair graph $\mathcal{G}$, we also consider the transition possibility of mobile devices to move from one position to the next one. Thus, we define the following transition probability as the edge weight.

\begin{definition}[Edge Weight] For a directed edge ${g_{i,j}\rightarrow g_{i+1,j'}}$ from a candidate grid $g_{i,j}$ to the next one $g_{i+1,j'}$ within a path of the repair graph $\mathcal{G}$. The edge weight is computed as follows.
\begin{equation}\scriptsize
	W_{g_{i,j}\to g_{i+1,j'}} \propto \frac{\cos\theta}{d(g_{i,j},g_{i+1,j'})}
    \label{edge_weight}
\end{equation}
\end{definition}

In the equation above, $d(g_{i,j},g_{i+1,j'})$ is the Euclidean distance between $g_{i,j}$ and $g_{i+1,j'}$, and $\theta$ is the angle between the two edges ${g_{i-1,j''}\rightarrow g_{i,j}}$ and ${g_{i,j}\rightarrow g_{i+1,j'}}$. The intuition to compute the edge weight $W_{g_{i,j}\to g_{i+1,j'}}$ is as follows. When one mobile device is walking or driving on a road, it is not likely to change the direction very frequently, and the physical distance between two neighbouring vertices (i.e., two neighboring locations) should not be very far away.

\begin{figure}
	\centering
	\includegraphics[totalheight=1.2in]{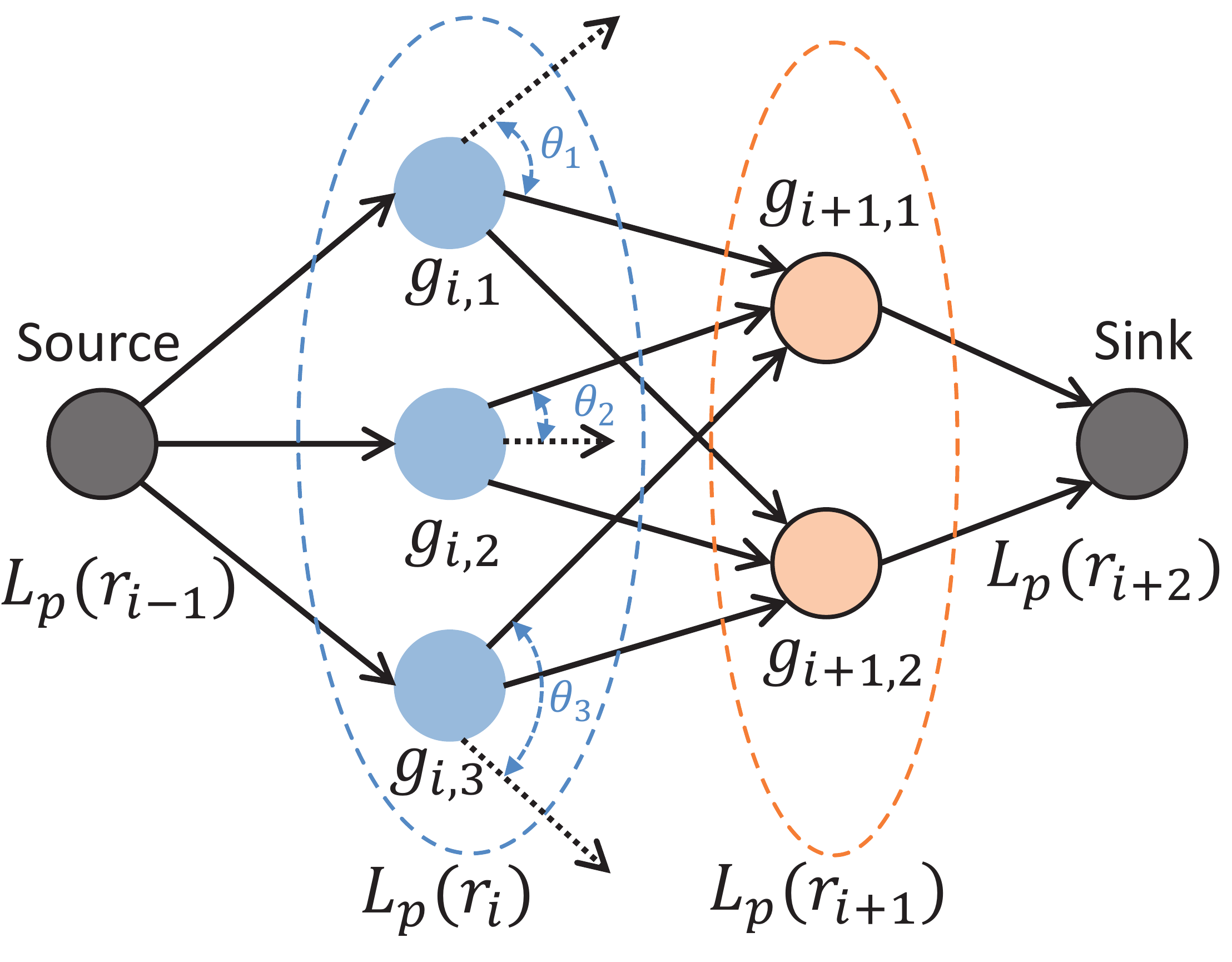}
	\caption{Example of a Repair Graph and Illustration of three angles: $\theta_1$ between the edges $L_{p}(r_{i-1})\rightarrow g_{i,1}$ and $g_{i,1}\rightarrow g_{i+1,1}$, $\theta_2$ between the edges $L_{p}(r_{i-1})\rightarrow g_{i,2}$ and $g_{i,2}\rightarrow g_{i+1,1}$, and $\theta_3$ between the edges $L_{p}(r_{i-1})\rightarrow g_{i,3}$ and $g_{i,3}\rightarrow g_{i+1,1}$} \label{fig:direction}
\end{figure}

\begin{example}
Figure \ref{fig:direction} illustrates an example repair graph $\mathcal{G}$, where the source and sink are normal. For the two flawed locations $L_p(r_i)$ and $L_p(r_{i+1})$, we have 3 candidate grids $g_{i,1}, g_{i,2}, g_{i,3}$ and 2 candidate grids $g_{i+1,1}, g_{i+1,2}$, respectively, and thus totally have 6 paths from source $g_{r_{i-1}}$ to sink $g_{r_{i+2}}$. Among the six paths, we choose one path with the maximal joint probability. The candidate grids within the selected path are then used to repair the flawed locations $L_p(r_{i})$ and $L_p(r_{i+1})$, respectively. In addition, this figure gives an example of three angles $\theta_1,\theta_2$ and $\theta_3$ between repair graph edges.

\end{example}

Until now, in the repair graph $\mathcal{G}$, each vertex $g_{i,j}$ is with a weight $W_{g_{i,j}}$ and each edge $g_{i,j}\rightarrow g_{i+1,j'}$ is with a weight $W_{g_{i,j}\to g_{i+1,j'}}$. Our task is to find a path from the source to sink, such that the found path is with the largest joint probability. Consider a path $\omega = g_1\rightarrow...\rightarrow g_{|\omega|}$ that traverses $|\omega|$ vertices from $g_1$ to $g_{|\omega|}$, we compute the joint probability of $\omega$.

\begin{equation}\scriptsize
P(\omega) = W_{g_1} \cdot W_{g_{1}\to g_{2}}\cdot W_{g_2} ... W_{g_{|\Omega|-1}} \cdot W_{g_{|\Omega|-1} \to g_{|\Omega|}} \cdot W_{g_{|\omega|}}
\end{equation}\label{eq:joint_pro}

%Among all possible paths from the source to sink in the graph $\mathcal{G}$, we find one path $\omega \in \Omega$ with the maximal joint probability $P(\omega)$.

\subsection{Algorithm Detail}

\begin{algorithm}\scriptsize
%  \SetKwProg{Fn}{Function}{}{}
  \SetKwFunction{FGetCand}{GetCandidateGrids}
  \SetKwFunction{FRepairSeq}{RepairGridSequence}
    \caption{DP-based Repair Algorithm}\label{alg:dynamic}
    \KwIn{$\mathcal{G}$ Repair Graph, $\mathcal{G}_{vertex}(\cdot)$ vertex weight, $\mathcal{G}_{edge}(\cdot)$ edge weight}
    \KwOut{$rSeq$ the path with the max. joint probability}

    {
    $JP[] \leftarrow$ the highest joint probability so far\;
    $par[]\leftarrow$ parent nodes of current candidates\;
    %$V \leftarrow$ vertices in $\mathcal{G}$\;
    $V_{parent}\leftarrow$ those vertices in $\mathcal{G}$ without parents\;
   \lForEach {$vp\in V_{parent}$}
   {
        $JP[vp]\leftarrow$ $\mathcal{G}_{vertex}(vp)$ 
    }
    \While{$V_{parent}$ still has children in $\mathcal{G}$}
    {
        $max$=$-\infty$, $V_{child}\leftarrow$ child vertices of $V_{parent}$ in $\mathcal{G}$\;
        \ForEach{$vc\in V_{child}$}
        {
            \ForEach{$vp\in V_{parent}$}{
                \lIf{$|V_{child}| == 1$}
                {
                   $tmp$=$JP[vp]$
                }
                \lElse
                {
                    $tmp$=$JP[vp] \times \mathcal{G}_{vertex}(vp) \times \mathcal{G}_{edge}(vp \to vc)$
                }
            \lIf{$tmp>max$}
            {
                $max=tmp;par[vc]=vp$
            }
            $JP[vc]$=$max$\;

            }

        }
        $V_{parent}\leftarrow V_{child}$\;
    }
    
    Initialize $rSeq$ as an empty list\;
    $c$=$argMax_{v}(JP[v])$, add $R$ to $rSeq$\;
    \lWhile{$par[c] \neq \emptyset$ }{
        add $par[c]$ to $rSeq$, $c \leftarrow par[c]$
    }
    \Return $rSeq$\;
  }

\end{algorithm}

Algorithm \ref{alg:dynamic} outlines the sequence-based repair via a dynamic programming method. It requires an input repair graph $\mathcal{G}$ and generates a trajectory or equivalently a path of selected candidate positions having the maximal joint probability. In general, finding such a path in a repair graph is NP-hard. Thus, we design an efficient path planning algorithm. The planning algorithm first finds the vertices $V_{parent}$ having no parent (line 3) after the initiation of two variables $JP[]$ (joint probability) and $par[]$ in lines 1-2. Next, the \textbf{loop} in lines 5-13 visits the remaining vertices level by level in the repair graph $\mathcal{G}$ by a Breadth-First Search (BFS) style. The $JP[]$ maintains the largest joint probabilities from sink to the current vertices so far. Thus, when the edges from $vp\in V_{parent}$ to $vc\in V_{child}$ are considered, we are interested in the maximal joint probability $JP[vp] \times \mathcal{G}_{vertex}(vp) \times \mathcal{G}_{edge}(vp\rightarrow vc)$, where $\mathcal{G}_{vertex}(vp)$ denotes the vertex weight of $vp$ and $\mathcal{G}_{edge}(vp \rightarrow vc)$ denotes the weight of the edge $vp \rightarrow vc$. Such maximal product is again maintained by a new item $JP[vp]$. Meanwhile the item $par[vc]$ with respect to $vc$ maintains the parent vertex $vp$. Once $V_{parent}$ has no child, the entire graph $\mathcal{G}$ has been visited and the sink has been reached. Thus, the algorithm breaks the \textbf{loop}. Now we simply find the item $v$ in $JP[]$ leading to the maximal $JP[v]$ (line 15). By reversely tracking the parent $par[c]$ of such found item $v$ (line 16), we can return a sequence of desirable candidates leading to the maximal $JP[v]$.

The running time of Algorithm \ref{alg:dynamic} heavily depends upon the path planning, especially the two \textbf{loops} in lines 8-13. Suppose the repair graph $\mathcal{G}$ has $n$ neighbouring flawed locations, i.e., $n$ levels from source to sink, and each level has $k$ candidates. Thus, the running time of the path planning part is $O(k^n)$. Note that the previous work \cite{LouZZXWH09} does not adopt any detection algorithm. Thus, when given an entire sequence of continuous grids (no matter flawed or not), the running time of \cite{LouZZXWH09} is $O(k^N)$, where $N$ is the size of the entire sequence. Thus, our repair algorithm can significantly improve the efficiency from $O(k^N)$ to $O(k^n)$ in particular due to $N \gg n$.

%% file: 05-evaluate.tex
\section{Evaluation}\label{s:evaluation}
In this section, we evaluate our approach \textsf{RLoc} in terms of three aspects: the overall localization accuracy after \textsf{RLoc} is applied to correct flawed positions, the performance of the proposed detection and repair algorithms, and sensitivity study of \textsf{RLoc} to key parameters.

\subsection{Experimental Setting}\label{sec:experimental setting}
\textbf{Data sets}: In Table \ref{tab:dataset},  we use totally three datasets: two collected from the rural Jiading district of North-west Shanghai, and one from the urban Xuhui district in the core center Shanghai (The physical distance between the two districts is around 31 km).

\begin{table}[!hbp]
%\small
\scriptsize
\caption{Statistics of Used Data Sets }\label{tab:dataset}
\centering
\begin{tabular}{|l|cc|cc|c|}
\hline
\multirow{2}{*}{} & \multicolumn{2}{c|}{Jiading-Campus} & \multicolumn{2}{c|}{Xuhui} & Jiading-Rural \\ \cline{2-6}
                  &2G &4G &2G &4G &4G \\
\hline
\hline
Num. of IMSIs   &7 &4&4&3&5967 \\%\hline
Num. of samples &20324 & 14218 & 24570 & 16905 & 150288\\%\hline
Sampling Period (sec) & 2$\thicksim$3 &2$\thicksim$3    & 1&1 &10$\thicksim$11 \\%\hline
Density of Serving Stations & 25.85 & 29.43  & 28.18 & 38.76 &24.92\\%\hline
Num. of Serving Stations& 61 & 44  & 21 & 16 &508\\%\hline
 Coverage Area ($km^2$)& \multicolumn{2}{c|}{1.64*1.44}    &\multicolumn{2}{c|}{1.32*0.43 }   & 4.46*4.57\\\hline
\end{tabular}

\end{table}

\begin{itemize}
    \item Jiading-Campus: This dataset, collected by our developed Android App, contains MR samples collected from 2G GSM and 4G LTE networks in a university campus that is located within the rural Jiading area. When students holding mobile devices installed with the App are moving around outdoor campus roads, the App then collects MR samples and current GPS coordinates. Table \ref{tab:mr} shows the data format of this dataset. 
    Note that, probably due to the limitation of Android API and policy rules of backend system configuration with respect to Telco networks, the identifiers (RNCID\_2$\sim$7 and CellID\_2$\sim$7) of non-serving base stations are null values, though the associated RSSI measurements could be collected in 4G MR samples.
    \item Xuhui: This dataset contains 2G and 4G samples collected on several main roads. As mentioned in Section \ref{sec:bkmrdata}, the data formats of MR samples collected by frontend Android APIs and backend operators may differ. For example, the backend 2G samples contain the signal measurements such as RxLev (= RSSI), ARFCN (absolute radio-frequency channel number) and the backend 4G MR samples contain the identifiers of all connected base stations and the associated signal measurements such as RSSI, RSRP and RSRQ. The detail of these data formats refer to the previous work \cite{DBLP:conf/mdm/HuangRZLYZY17}.
    \item Jiading-Rural: This large dataset contains 4G LTE MR samples collected in a large rural area in Jiading. The sampling rate of this dataset is rather low, i.e., one sample for every 10$\thicksim$11 seconds, when compared with other datasets. This dataset follows the same data format as Xuhui dataset.
\end{itemize}

 Similar to the previous works NBL \cite{MargoliesBBDJUV17} and CCR \cite{ZhuLYZZGDRZ16}, we use GPS coordinates as the ground truth locations of MR samples. Since the collected GPS coordinates may contain noisy information, we exploit the map-matching technique \cite{HuangRZZYZ18} to mitigate the effect of noisy information. To protect user privacy, all IMSIs (International Mobile Subscriber Identity) in the used datasets have been anonymized.

\textbf{Counterparts and Data Division}: In Table \ref{tab:method}, we evaluate \textsf{RLoc} against four counterparts, including three  outdoor localization approaches: a Random Forest regressor-based approach CCR \cite{ZhuLYZZGDRZ16}, HMM-based localization approach \cite{DBLP:conf/IEEEcit/NiWTYS17} (for simplicity we rename this HMM-based approach as HLoc), and fingerprinting-based localization approach NBL \cite{MargoliesBBDJUV17}) and our previous data repair-based approach CRL \cite{ZhangRYZY17}. Note that HLoc \cite{DBLP:conf/IEEEcit/NiWTYS17} originally works only on 4G LTE data and requires the items of both TA (Timing Advance) and RSRP (Reference Signal Receiving Power). Since the MR samples in our used datasets do not contain the TA item and the MR samples in 2G datasets or frontend Android datasets do not contain the RSRP item, for fairness, our implementation of \cite{DBLP:conf/IEEEcit/NiWTYS17} has to remove the component regarding TA and then replace RSRP by RSSI.

These five approaches all require localization steps, and only two of them \textsf{RLoc} and \textsf{CRL} \cite{ZhangRYZY17} require the localization, detection and repair steps. Here, both \textsf{RLoc} and \textsf{CRL} \cite{ZhangRYZY17} use a RaF classifier-based localization model $\mathcal{L}$, whereas CCR \cite{ZhuLYZZGDRZ16} adopts a RaF regressor-based localization model. The input to the three Raf-based localization algorithms contain the features such as raw MR features (see Table \ref{tab:mr}), base station features (e.g., GPS coordinates of base stations) provided by Telco operators, and hand-made contextual features (e.g., the moving speed and direction \cite{ZhuLYZZGDRZ16}). Nevertheless, \textsf{RLoc} and \textsf{CRL} differ in terms of the used detection and repair algorithms: \textsf{RLoc} exploits the sequenced-based approach, and yet \textsf{CRL} the single-point-based approach.

We give the training and testing data of the five approaches as follows. \textsf{CCR}, \textsf{HLoc} and \textsf{NBL} do not require detection and repair algorithms. We thus assign the entire datasets $\mathbb{D}$ and $\mathfrak{D}$ to train and test a localization model for them, respectively. Instead, besides the localization model $\mathcal{L}$, \textsf{RLoc} and \textsf{CRL} require detection and repair algorithms. Thus, we assign the subset $\mathbb{D}_L$ as the training dataset for $\mathcal{L}$ and the subset $\mathbb{D}_C$ as the training data for the detection/repair algorithms. In this way, the same training dataset $\mathbb{D}$ is assigned to all five approaches, and we do not assign extra more samples to train the localization/detection/repair algorithms for \textsf{RLoc} and \textsf{CRL}. Thus, our data assignment guarantees evaluation fairness for five approaches.

For the proportion of MR samples assigned for $\mathbb{D}_L$, $\mathbb{D}_C$, and $\mathfrak{D}$, we divide the samples in each MR dataset into three disjoint parts (see Table \ref{tab:method}). Specifically, to avoid over-fitting, we adopt the 10-fold cross validation \cite{DBLP:conf/ijcai/Kohavi95} and randomly choose 80\% samples for $\mathbb{D}$ and 20\% for $\mathfrak{D}$. Among the samples in $\mathbb{D}$, we further randomly assign 62.5\% samples for $\mathbb{D}_L$ and 37.5\% samples for $\mathbb{D}_C$. We implement the five approaches with Python and evaluate them on a Linux workstation with Intel(R) Xeon(R) CPU E5-2620 v3 @2.40GHz and 64 GB memory.

\begin{table}[h]
%\small
\scriptsize
\caption{Counterparts and Used Data Sets ($\mathcal{L}$: Localization Model,  $\mathcal{C}$: Confidence-based Detection)}\label{tab:method}
\centering
\begin{tabular}{|l|l|l|l|c|}
\hline
	& $\mathcal{L}$, Training Data& $\mathcal{C}$, Training  Data& Repair& Testing Data\\
\hline
\hline
RLoc &RaF classifier, $\mathbb{D}_L$ & DA-HMM, $\mathbb{D}_C$  & DP&  $\mathfrak{D}$\\%\hline
CRL \cite{ZhangRYZY17} &RaF classifier, $\mathbb{D}_L$  & GBDT, $\mathbb{D}_C$ & Baseline& $\mathfrak{D}$\\%\hline%\hline
CCR \cite{ZhuLYZZGDRZ16}& RaF regression, $\mathbb{D}$ &  &   & $\mathfrak{D}$\\%\hline
HLoc \cite{DBLP:conf/IEEEcit/NiWTYS17} & HMM, $\mathbb{D}$ &  & & $\mathfrak{D}$\\%\hline
NBL \cite{MargoliesBBDJUV17} & Fingerprinting, $\mathbb{D}$ &  & & $\mathfrak{D}$\\\hline
\end{tabular}

\end{table}

\textbf{Performance Metrics and Key Parameters}: Firstly, we are interested in the metric of \emph{localization errors}. Specifically, for each MR sample in the testing dataset, we first predict its location and then compute the error by the Euclidean distance between the predicted location and ground truth. We use the cumulative distribution function (CDF) of errors in the testing data set to evaluate the localization performance, and the key indicators are \emph{mean}, \emph{median} (top-50\%), \emph{top-67}\%, \emph{top-90}\% and \emph{top-95}\% errors.

Secondly, we evaluate the confidence-based detection algorithm by three metrics: \emph{Precision} $M_p =  \frac{|D \sqcap G|}{|D|}$, \emph{Recall} $M_r = \frac{|D \sqcap G|}{|G|}$, and $F$-\emph{score} $M_f =\frac{2*M_p*M_r}{M_p+M_r}$, where $D$ is the set of detected flawed MR samples and $G$ is the set of the ground truth. Here, the ground truth of flawed samples can be achieved if the criteria of confidence levels $||L_p(r)-L_t(r)||\ge \tau$ is met.

Thirdly, we define the \emph{repair accuracy} of a repair algorithm by $r_{\alpha} = \frac{|R|}{|D|}$, where $R$ is the set of those correctly repaired samples among all detected flawed samples $D$. Next, we are interested in how much localization error is reduced after the repair algorithm is applied. Thus, we measure the \emph{repair ratios} of the reduced errors by the repair algorithm. Specifically, given the original median, 67\% and 95\% localization errors (denoted by $E^{d}$, $E^{s}$ and $E^{l}$) before the repair is applied and those (denoted by  $E_{r}^{d}$, $E_{r}^{s}$ and $E_{r}^{l}$) after the repair is applied, we define three {repair ratios} $I_{d} =\frac{E^{d}-E_{r}^{d}}{E^{d}}$, $I_{s} =\frac{E^{s}-E_{r}^{s}}{E^{s}}$, and $I_{l}=\frac{E^{l}-E_{r}^{l}}{E^{l}}$. Moreover, to measure the quality of a candidate selection approach, among the selected candidate set $C$, we are interested in \emph{1}) the precision $p_c$, i.e., the proportion of ground truth grids that appear within the candidate set, and \emph{2}) the number $|C|$ of selected candidates per flawed sample. Intuitively, we will select a small number $|C|$ of candidates and yet repair flawed positions with a high repair precision $p_c$.

Finally, Table \ref{tab:param} lists the mainly used parameters. Since we use the error threshold $\tau$ in Section \ref{s:prob} and Section \ref{sec:conf1} to determine flawed MR samples and confidence levels, we thus set $\tau$ by a relatively high localization error and vary it from top 70\% error to top 90\% error. In addition, the threshold $\gamma$ in Section \ref{sec:aemission} determines whether or not $|\mathbb{D}_C(v_k^{bs})|$ is a trivial sample size. Thus, we empirically tune it by relatively small values from 1 to 15. In terms of similarity threshold $\varepsilon$ in Section \ref{sec:aemission}, it determines whether the two MR base station observations $v_k^{bs}$ and $v_{k'}^{bs}$ are similar. Since we measure the similarity by Jaccard coefficient within the range $[0.0, 1.0]$, we tune $\varepsilon$ from a small value 0.25 to the maximal one 1.0 in order to find sufficient samples. Finally, the threshold $\xi$ in Section \ref{sec:candpos} is used to select candidate positions, we vary $\xi$ from a relatively high value 0.6 to the maximal one 1.0 to guarantee the quality of selected candidate positions. We use default values if without special mention, and vary their values within the allowable range for sensitivity study.
\begin{table}[!hbp]
		\caption{Parameters and Default Values}\label{tab:param}
		\scriptsize
	\centering
	\begin{tabular}{|l|c|c|}
		\hline
		&Allowable Range &Default Val.\\
		\hline
		%\hline
		%Grid Size (meters)& $10 - 40$ &30 \\\hline
        Error Threshold $\tau$ in Sections \ref{s:prob} and \ref{sec:conf1}  & 70\% error $-$ 90\% error &80\% error \\%\hline
        Sample Size Threshold $\gamma$ in Section \ref{sec:aemission} & $1-15$ &5 \\%\hline
        Similarity Threshold $\varepsilon$ in Section \ref{sec:aemission} &$0.25-1.0$ & 0.5 \\%\hline
        Similarity Threshold $\xi$ in Section \ref{sec:candpos}& $0.6-1.0$ & 0.7\\\hline
	\end{tabular}

\end{table}

\subsection{Localization Performance}

\begin{figure*}[th]
%\hspace{-8ex}
\begin{center}
\begin{tabular}{c c c c c}
\begin{minipage}[t]{0.19\linewidth}
\begin{center}
\centerline{\includegraphics[totalheight=1.1in]{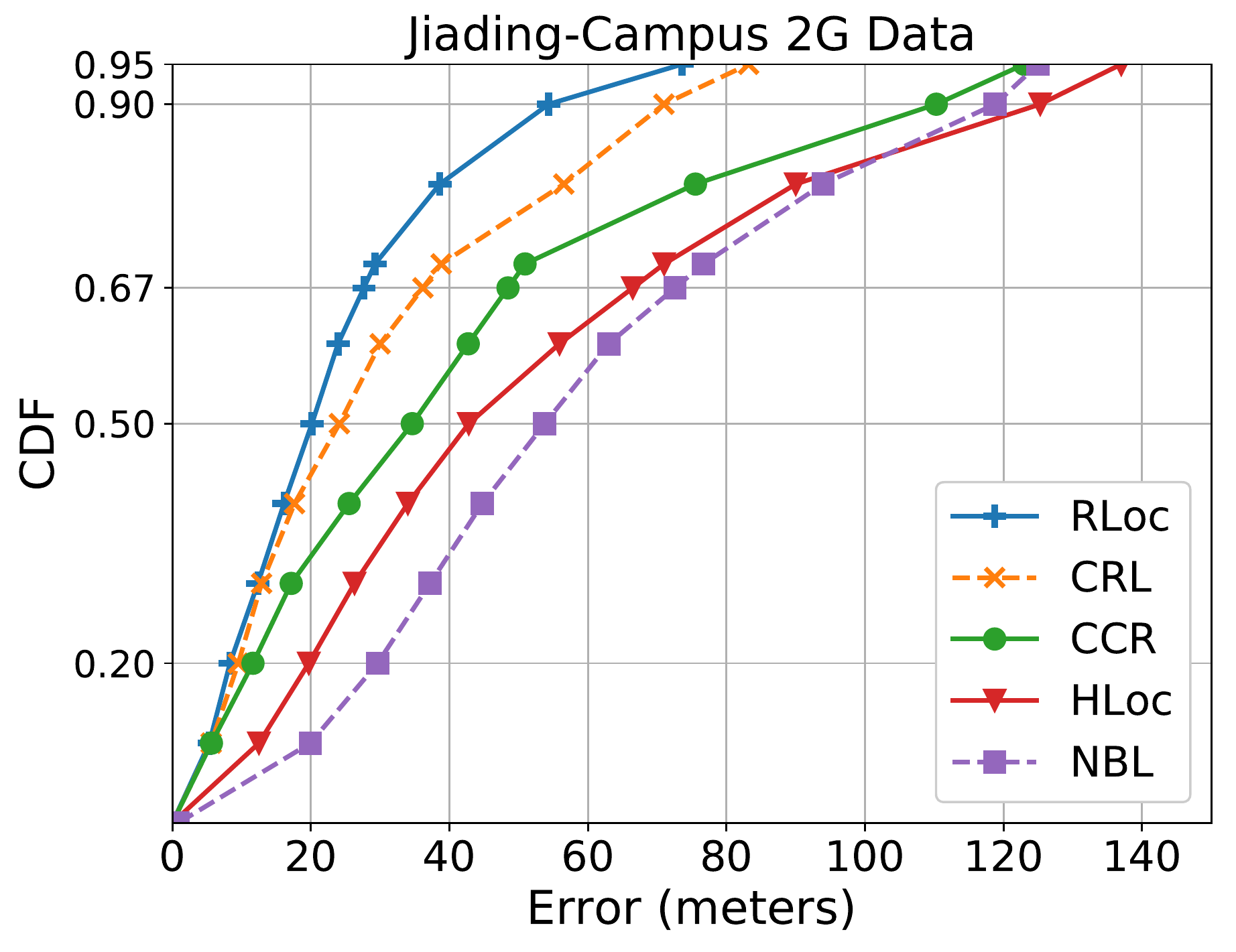}}
\end{center}
\end{minipage}
&
\begin{minipage}[t]{0.19\linewidth}
\begin{center}
\centerline{\includegraphics[totalheight=1.1in]{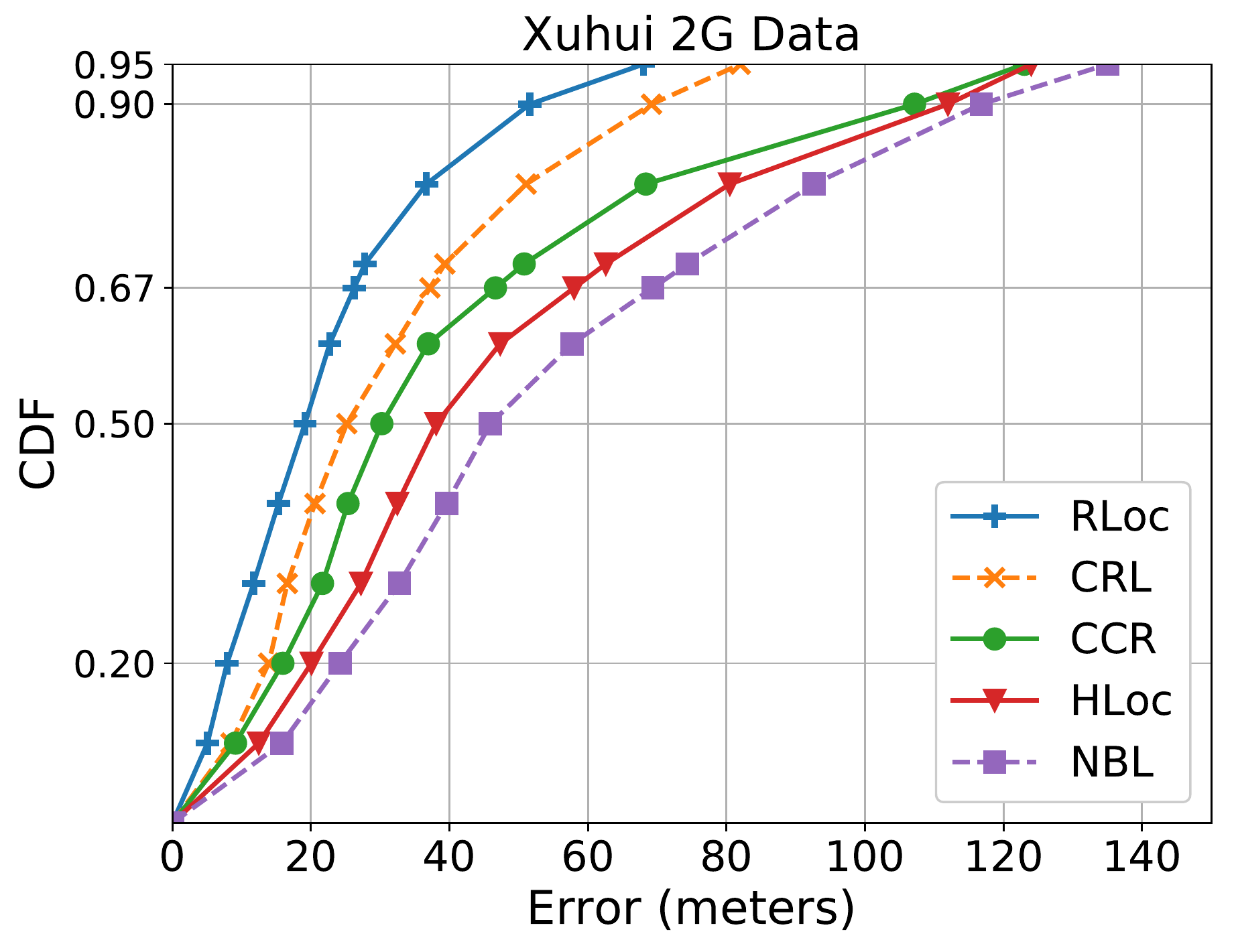}}
\end{center}
\end{minipage}
&
\begin{minipage}[t]{0.19\linewidth}
\begin{center}
\centerline{\includegraphics[totalheight=1.1in]{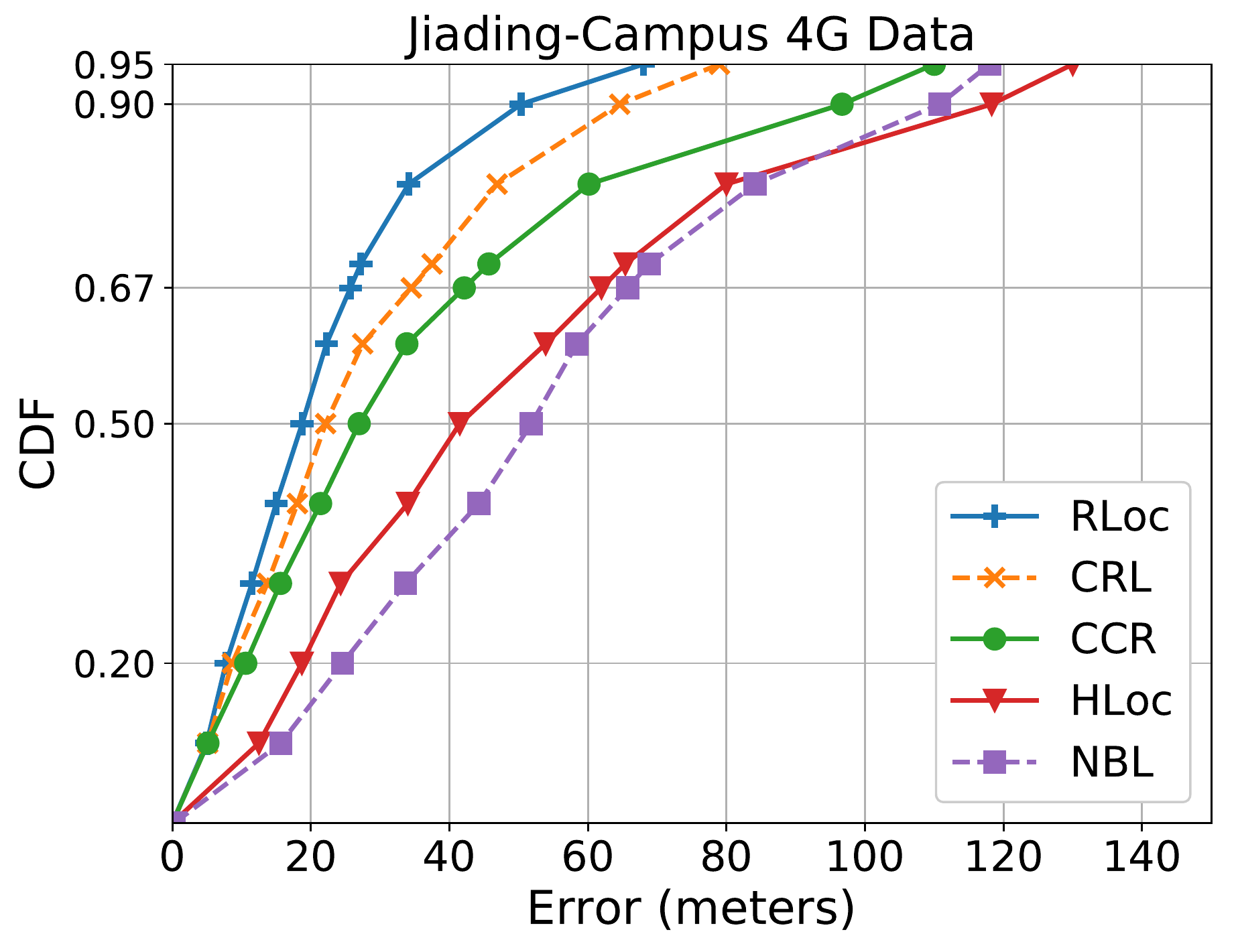}}
\end{center}
\end{minipage}
&
\begin{minipage}[t]{0.19\linewidth}
\begin{center}
\centerline{\includegraphics[totalheight=1.1in]{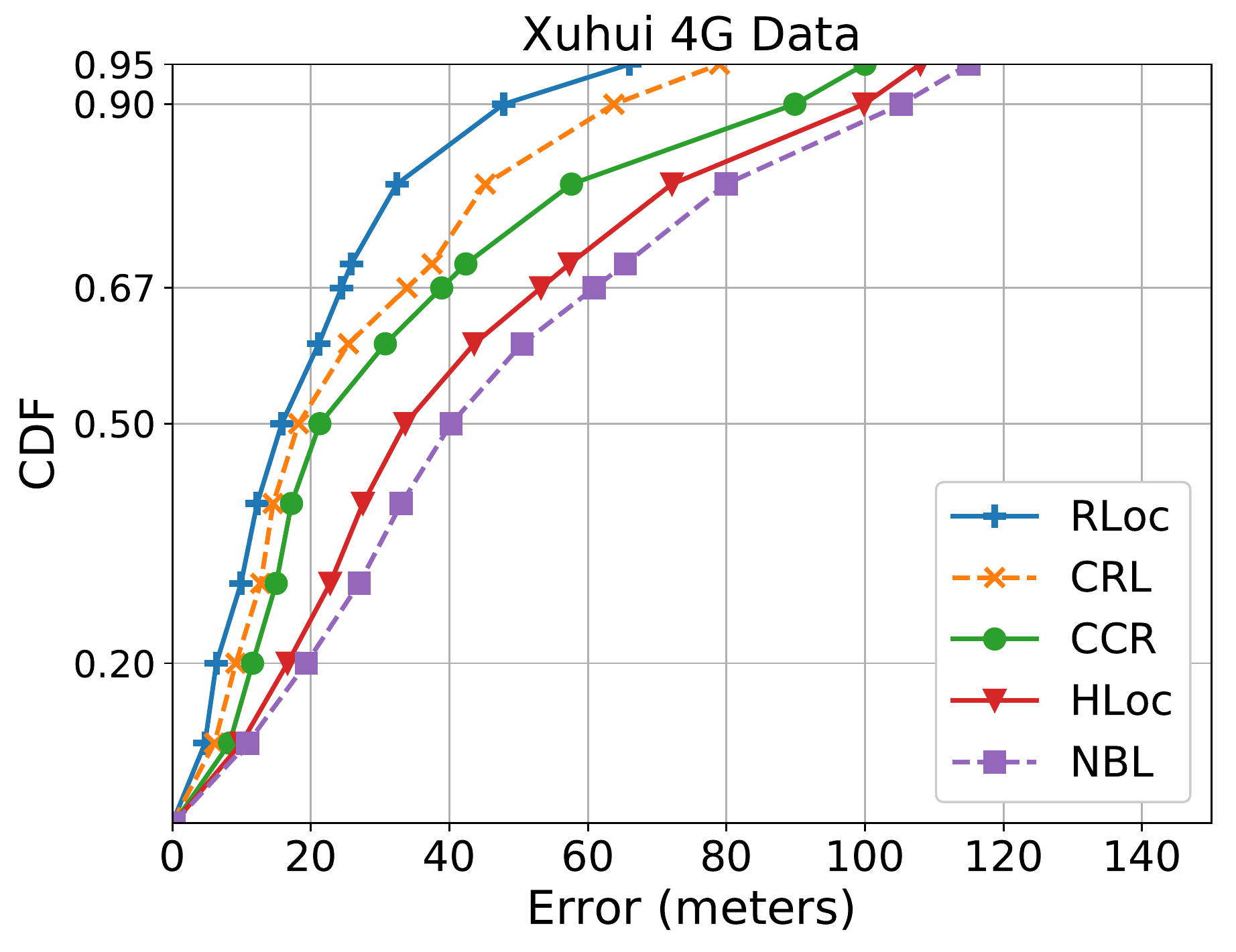}}
\end{center}
\end{minipage}
&
\begin{minipage}[t]{0.19\linewidth}
\begin{center}
\centerline{\includegraphics[totalheight=1.1in]{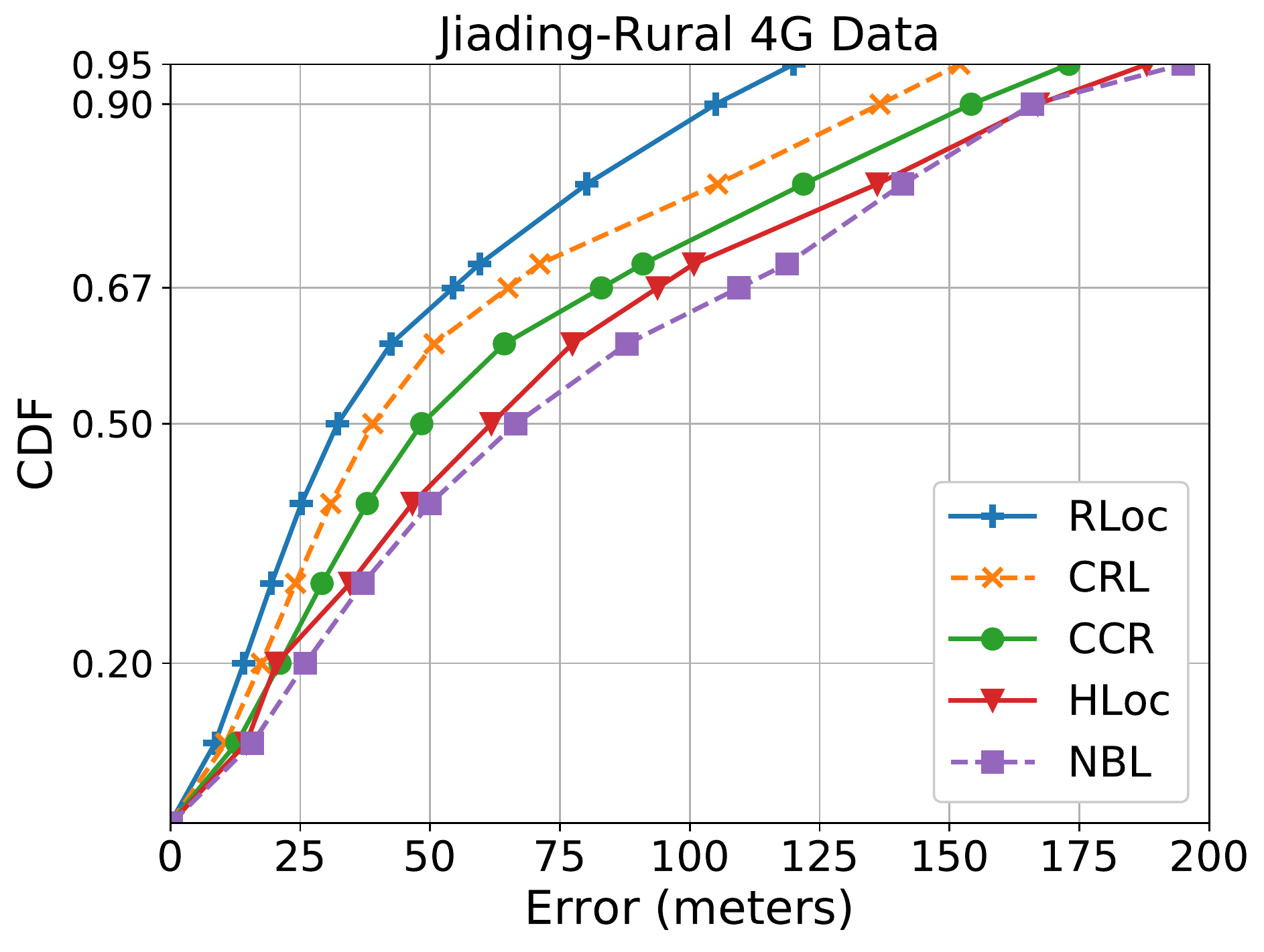}}
\end{center}
\end{minipage}
\end{tabular}\vspace{-4ex}
\caption{Localization Errors of Five Algorithms on 5 Used Data Sets: Jiading-Campus 2G, Xuhui 2G, Jiading-Campus 4G, Xuhui 4G and Jiading-Rural 4G Datasets (from left to right).}
\label{exp:cdf}
\end{center}
\end{figure*}
Figure \ref{exp:cdf} gives the localization errors of five approaches, where the $x$-axis is the location error (meters), and $y$-axis is the (empirical) Cumulative Distribution Function -- CDF of localization errors. From this figure, we have the following findings.
\begin{itemize}
\item Firstly, the proposed \textsf{RLoc} greatly outperforms the four counterparts. For example, in the Jiading 4G data set, the median errors of \textsf{RLoc}, \textsf{CRL}, \textsf{CCR} , \textsf{HLoc} and \textsf{NBL} are 32.20, 38.98, 48.40, 61.83 and 66.51 meters, respectively. \textsf{RLoc} reduces the median error by 17.4\% compared with the state-of-the-art (\textsf{CRL}). These numbers indicate that \textsf{RLoc} can correct flawed locations for the best results. Moreover, among the two repair-based localization approaches, \textsf{RLoc} has smaller errors than \textsf{CCR}, mainly due to the sequence-based detection and repair.
\item Secondly, in both 2G and 4G data sets, the median errors of \textsf{HLoc} and \textsf{NBL} are much greater than the three other approaches. These numbers indicate that the fingerprinting-based and HMM-based approaches cannot compete the RaF-based approaches such as \textsf{CCR}. It is mainly because these RaF-based approaches leverage the rich engineered features of both MR samples and base stations (e.g., GPS coordinates of base stations). Instead, \textsf{NBL} uses only the MR features (e.g., signal strength) but not the features of base stations. Moreover, \textsf{HLoc} essentially employs a static HMM model and can not capture adaptive transition probability between spatial cell grids that is instead the main focus of the proposed DA-HMM. In addition, both \textsf{RLoc} and \textsf{CRL} lead to better performance than non-repair localization approach \textsf{CCR}, especially with the 90\% and 95\% errors. This result indicates that the detection and repair algorithms work rather well to correct outlier flawed locations.
\item Thirdly, in Jiading-Campus and Xuhui datasets, all algorithms achieve better localization result in 4G data than the one in 2G data set. It is manly because 4G Telco networks typically deploy more dense base stations than 2G Telco networks. The 4G MR samples hence frequently contain much stronger Telco signal strength. In addition, the localization errors of all algorithms in Xuhui data sets are slightly lower than those in Jiading-Campus data sets. This is also due to the dense base stations deployed in the urban Xuhui area and sparse ones in the rural Jiading area. Note that the localization errors of Jiading-Rural 4G data set are higher than the other four data sets. It is mainly due to the smallest sampling rate among all datasets and more sparse base station density of Jiading area than the one of Xuhui area.
  \end{itemize}

Due to space limit, in the rest of this section, we evaluate the performance of \textsf{RLoc} on Jiading-Rural 4G LTE dataset (with the greatest amount of data samples on the largest area).

\subsection{Detection Performance}

\begin{figure*}[th]
	%\hspace{-8ex}
	\begin{center}
		\begin{tabular}{c c c c c}
			\begin{minipage}[t]{0.19\linewidth}
				\begin{center}
					\centerline{\includegraphics[totalheight=1.1in]{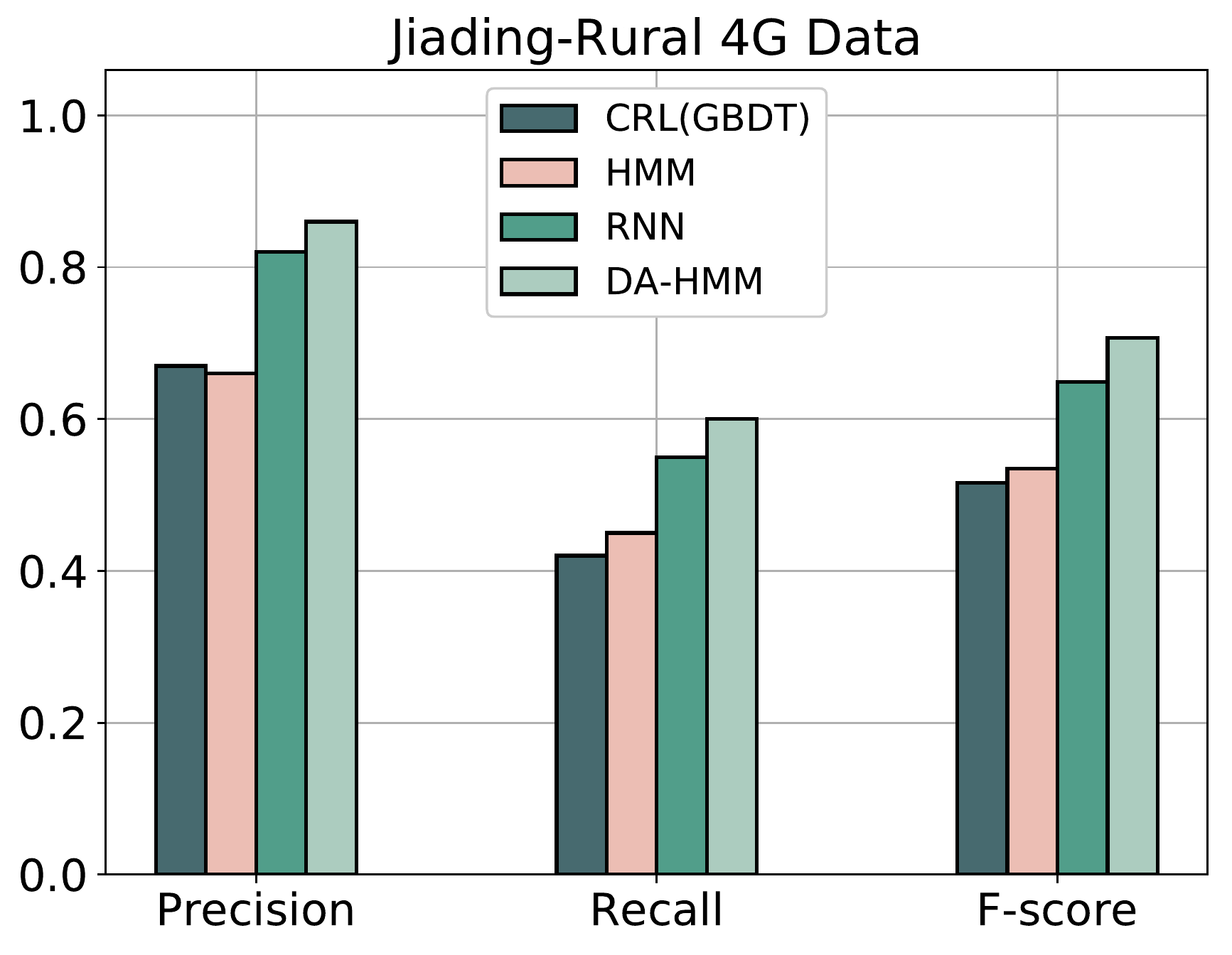}}
					%\centerline{(a)}
				\end{center}
           % \caption{Evaluation of Detection Algorithms.}
%\label{fig:detect}%\vspace{-2ex}
			\end{minipage}
			& %\hspace*{-2ex}
			\begin{minipage}[t]{0.19\linewidth}
				\begin{center}
					\centerline{\includegraphics[totalheight=1.1in]{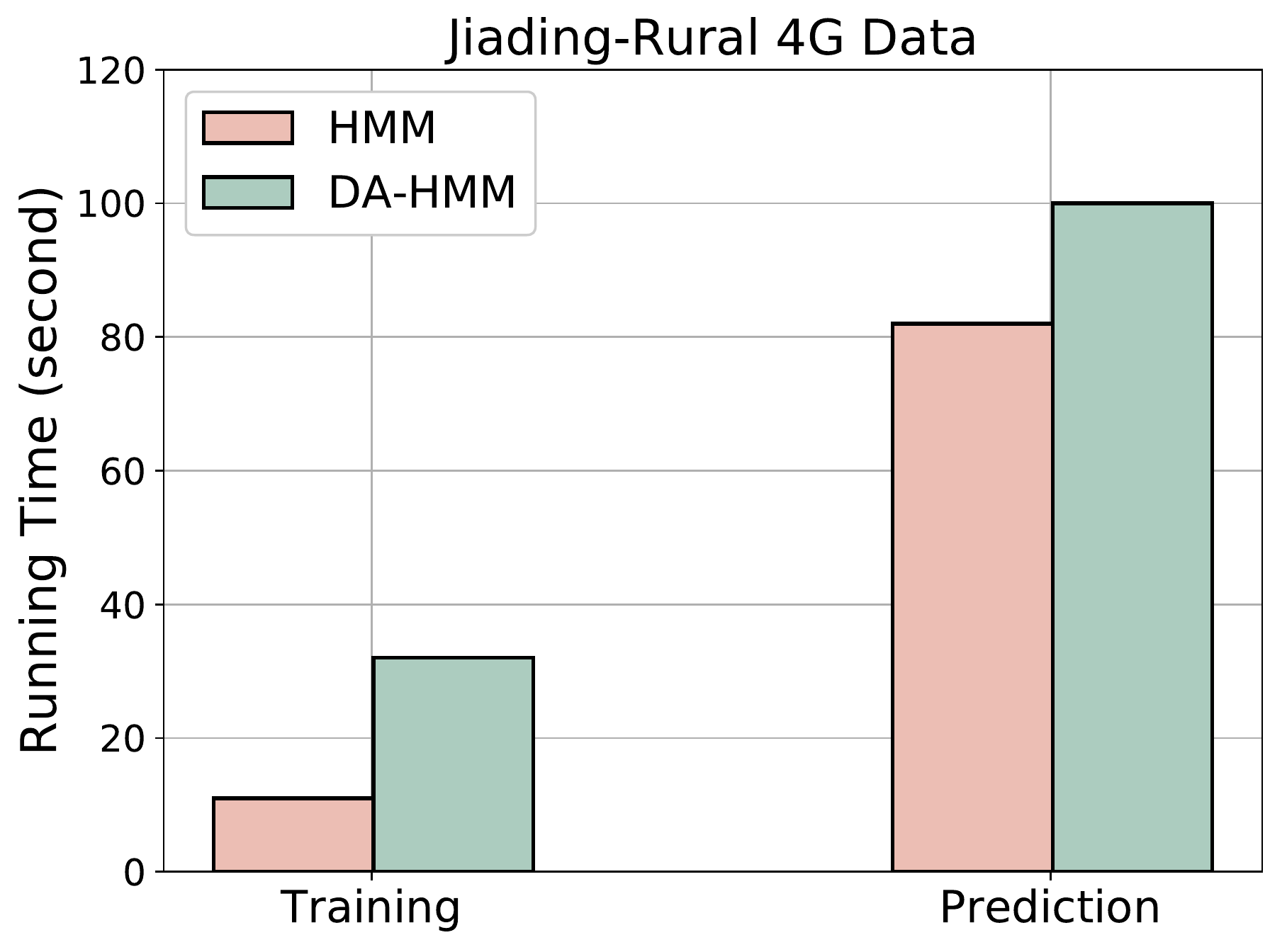}}
					%\centerline{(b)}
				\end{center}
              %  \caption{Running Time of HMM-based Algorithms.}
		     \label{exp:run_time}%
			\end{minipage}
            &
            \begin{minipage}[t]{0.19\linewidth}
				\begin{center}
					\centerline{\includegraphics[totalheight=1.1in]{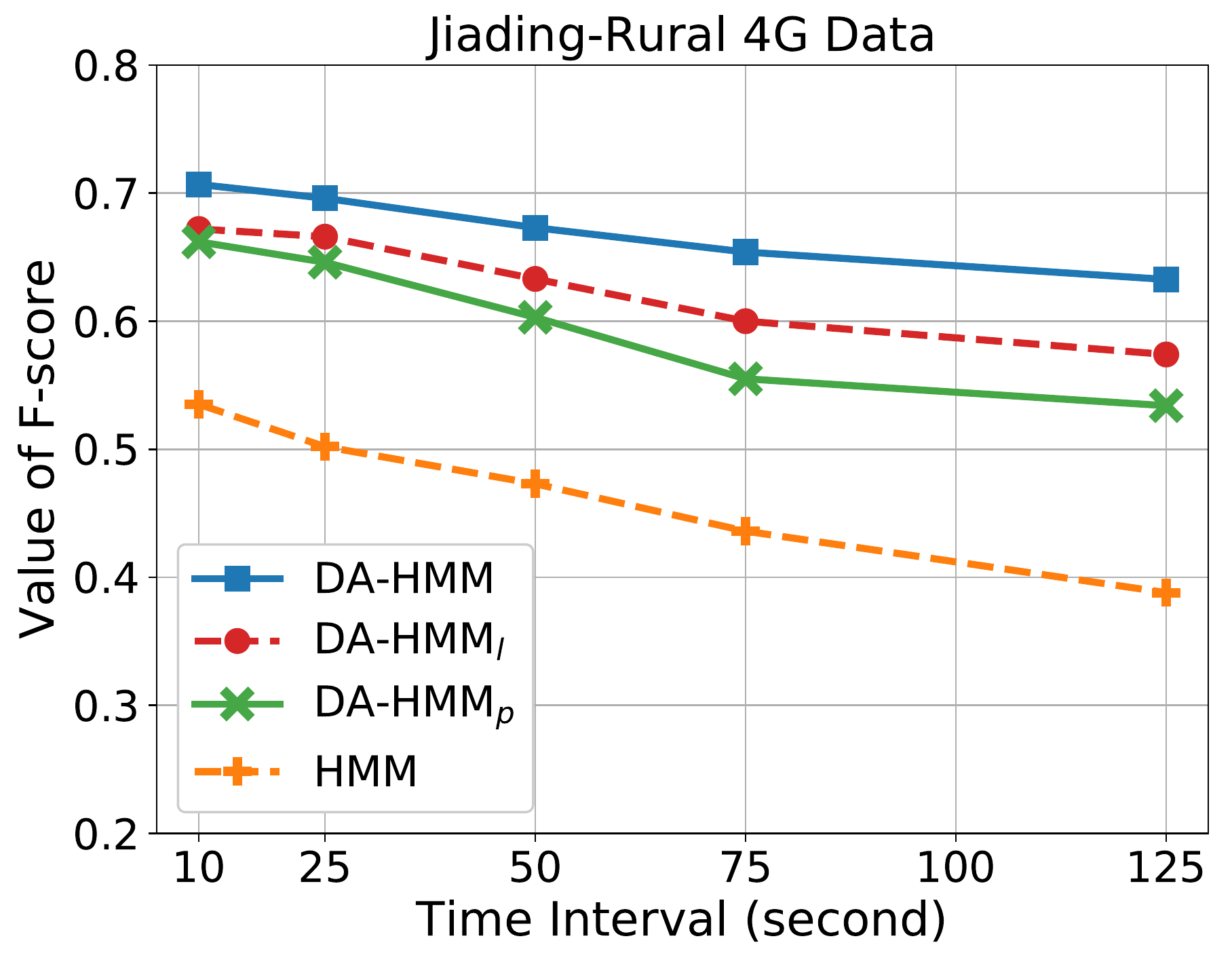}}
					%\centerline{(a)}
				\end{center}%\label{exp:vary_time_interval}
              %  \caption{Effect of Time Intervals on HMM-based Algorithms.}
		
			\end{minipage}
            &
            \begin{minipage}[t]{0.19\linewidth}
				\begin{center}
					\centerline{\includegraphics[totalheight=1.1in]{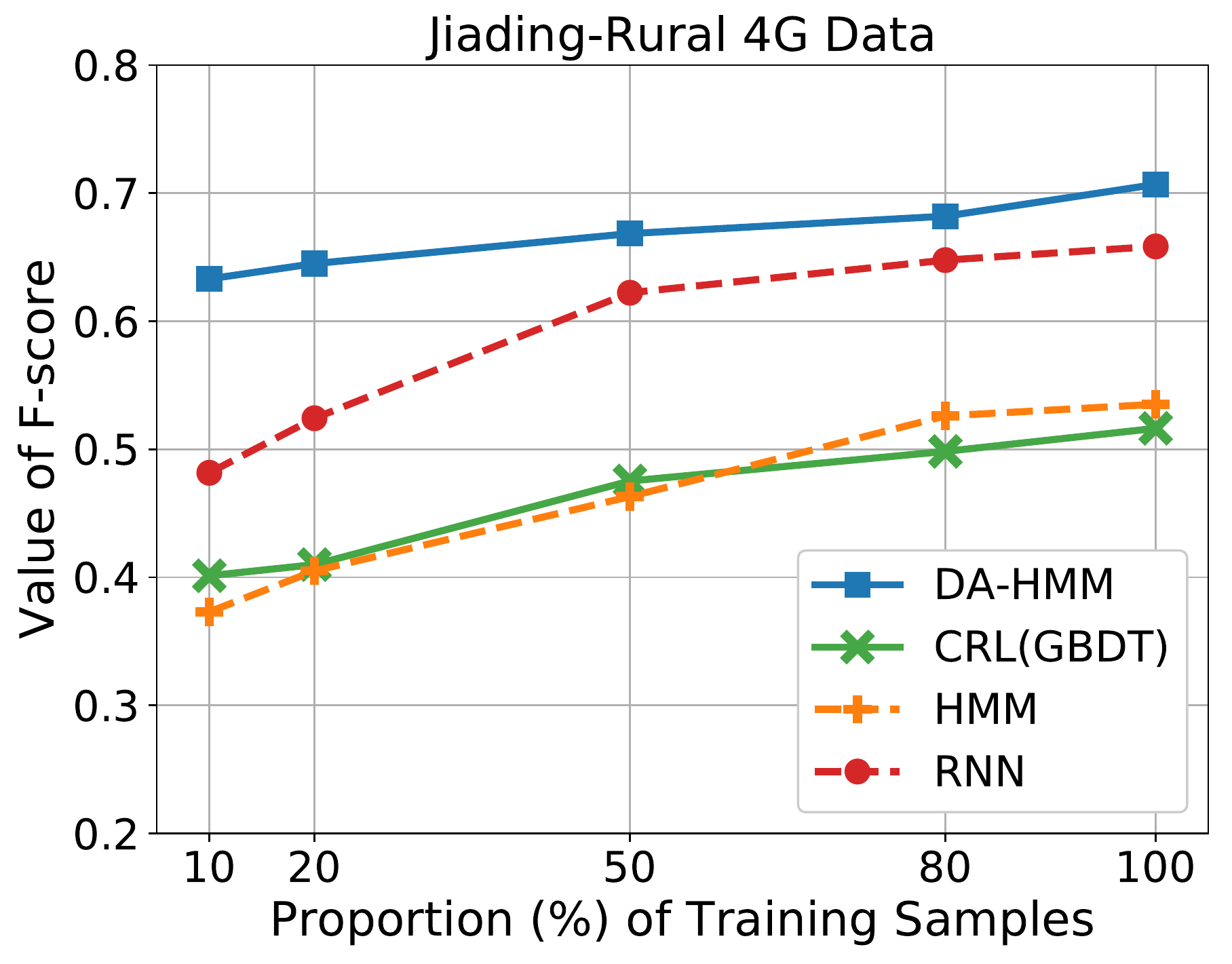}}
					%\centerline{(a)}
				\end{center}
                % \caption{Effect of Utilization Rate (\%) on HMM-based Detection Algorithms.}
		          %\label{exp:vary_training_sample}%\vspace{-2ex}
			\end{minipage}
            &
            \begin{minipage}[t]{0.19\linewidth}
				\begin{center}
					\centerline{\includegraphics[totalheight=1.1in]{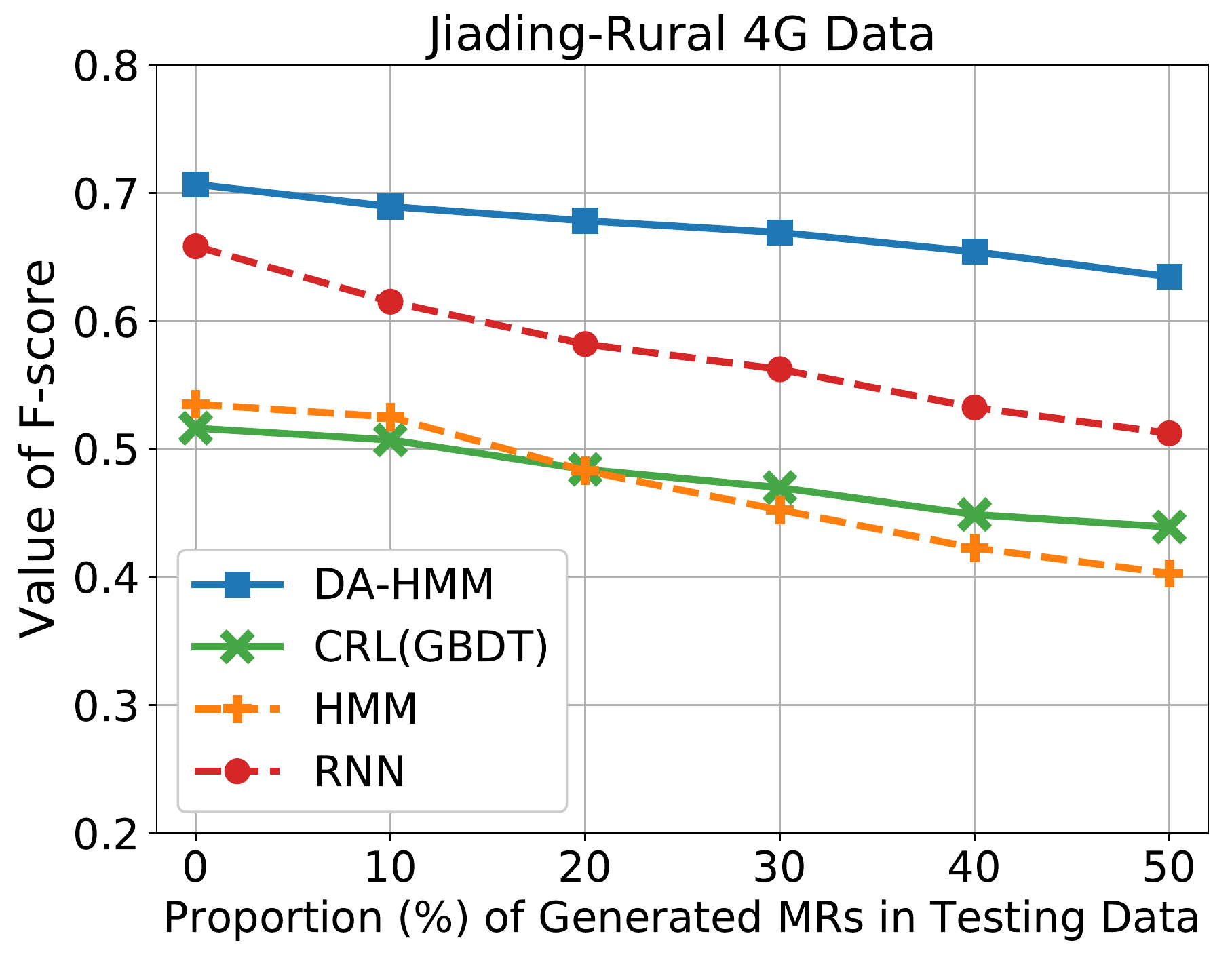}}
					%\centerline{(a)}
				\end{center}
			\end{minipage}
        \end{tabular}
	\end{center}\vspace{-5ex}
\caption{Detection Algorithm: (a-b) Effectiveness and Efficiency of Detection Approaches, (c-e) Effect of Time Intervals, Used Training Samples, and Generated MR samples in Testing Data (from left to right).}	\label{fig:detect}
\end{figure*}

In this section, we study the performance of four detection algorithms: the proposed DA-HMM model, static HMM model, the single-point-based GBDT classifier used by \textsf{CRL}, and a deep sequence model using the basic recurrent neural network (RNN).

Firstly, Figure \ref{fig:detect}(a) gives the precision, recall and F-score of the four approaches on the Jiading-Rural 4G data set. The static HMM (denoted as HMM) does not work well, and DA-HMM instead leads to the best result mainly due to the introduced adaptive probabilities $a_{i,j}^{\Delta}$ and $b_j^\gamma(k)$ to incorporate uncertain time-intervals and sample size. The RNN approach cannot compete DA-HMM, indicating that the DA-HMM model optimized by the adaptive probabilities could outperform the basic RNN model.

Secondly, Figure \ref{fig:detect}(b) plots the running time (used by training and testing phases) of the static HMM and DA-HMM approaches. Though DA-HMM requires around $3\times$ training time over the static HMM, the prediction time of DA-HMM is only $1.18\times $ of the static HMM. During the training phase, DA-HMM needs to estimate the parameters of two adaptive probabilities on top of the static HMM, thus the training time of DA-HMM is much higher than the one of static HMM. In terms of testing phase to infer the exact values of two adaptive probabilities, DA-HMM calculates the specific time intervals between neighbouring MR samples within testing sequence data. Since the remaining prediction steps of DA-HMM are consistent with static HMM, DA-HMM leads to slightly higher testing time cost.

Thirdly, we are interested in how DA-HMM performs on the MR samples with various neighboring time intervals. To this end, among the Jiading-Rural 4G data set, for each MR sequence, we randomly select some MR samples of the sequence to make sure that every timestamp difference between neighboring selected MR samples is no more than a certain value. Given these selected samples, we evaluate the proposed exponential regression-based DA-HMM (used for the adaptive state transition probability) against static HMM and two variants of DA-HMM using logistic regression and polynomial regression (denoted as DA-HMM$_{l}$ and DA-HMM$_{p}$, respectively). In Figure \ref{fig:detect}(c), we plot the F-score of the static HMM and three variants of DA-HMM (using the exponential, logistic and polynomial regression models). On the overall, a higher time interval means more sparse time sampling rate and thus worse detection performance. In addition, we note that the three DA-HMM approaches outperform the static HMM. It is mainly because the adaptive transition probability in DA-HMM can tackle the issue of various time intervals. In addition, among three regression algorithms, the exponential function leads to the best performance under various time intervals.

Fourthly, we are interested in the effect of the amount of used  training samples $\mathbb{D}_C$ on DA-HMM by tuning the adaptive emission probability, and thus vary the proportion of $\mathbb{D}_C$ from 10\% to 100\%. As shown in Figure \ref{fig:detect}(d), more training samples lead to higher F-score values for all four algorithms. In terms of the two HMM-based methods, the static HMM is more sensitive to the amount of training samples than DA-HMM, and whereas DA-HMM is adaptive to sparse training data. It makes sense because the design objective of the adaptive probabilities in DA-HMM is to overcome the issue of uncertain data sampling rate including sufficient and sparse data. In addition, since training a deep neural network RNN usually needs a large amount of training data, F-score of RNN drops rapidly when the proportion of $\mathbb{D}_C$ decreases from 50\% to 10\%. Finally, GBDT in general performs worst especially when training data is insufficient.

Finally, we study the generalization ability of DA-HMM by introducing a certain number of new testing MR samples in $\mathfrak{D}$. To this end, we follow a recent work \cite{ShokryTY18} to generate new MR samples by using the spatial and scan augmentation methods. By varying the proportion of these new samples from 0\% to 50\% in $\mathfrak{D}$, Figure \ref{fig:detect}(e) plots the F-scores of four approaches. As shown in this figure, more generated samples degrade the F-score of all detection algorithms. It is mainly because the new MR samples may not follow the same distribution of MR features (such as RSSI). Nevertheless, DA-HMM can still lead to competitive performance even if 50\% testing data are generated samples, and instead the static HMM and RNN-based methods are rather sensitive to the amount of generated samples than DA-HMM and GBDT. Note that this evaluation result differs from the one in the work \cite{ShokryTY18} which instead uses the new MR samples and original training MR samples ($\mathbb{D}$) together to train a localization model $\mathcal{L}$ for better localization accuracy. Yet in our experiment, we introduce generated samples to verify the generalization ability of DA-HMM (trained by $\mathbb{D}_C$ alone, but without any generated MR samples). Thus, it makes sense that a larger proportion of generated samples in $\mathfrak{D}$ could degrade the accuracy of DA-HMM.

\begin{figure}[th]
	%\hspace{-8ex}
	\begin{center}
		\begin{tabular}{c c }
			
            \begin{minipage}[t]{0.47\linewidth}
				\begin{center}
					\centerline{\includegraphics[totalheight=1.1in]{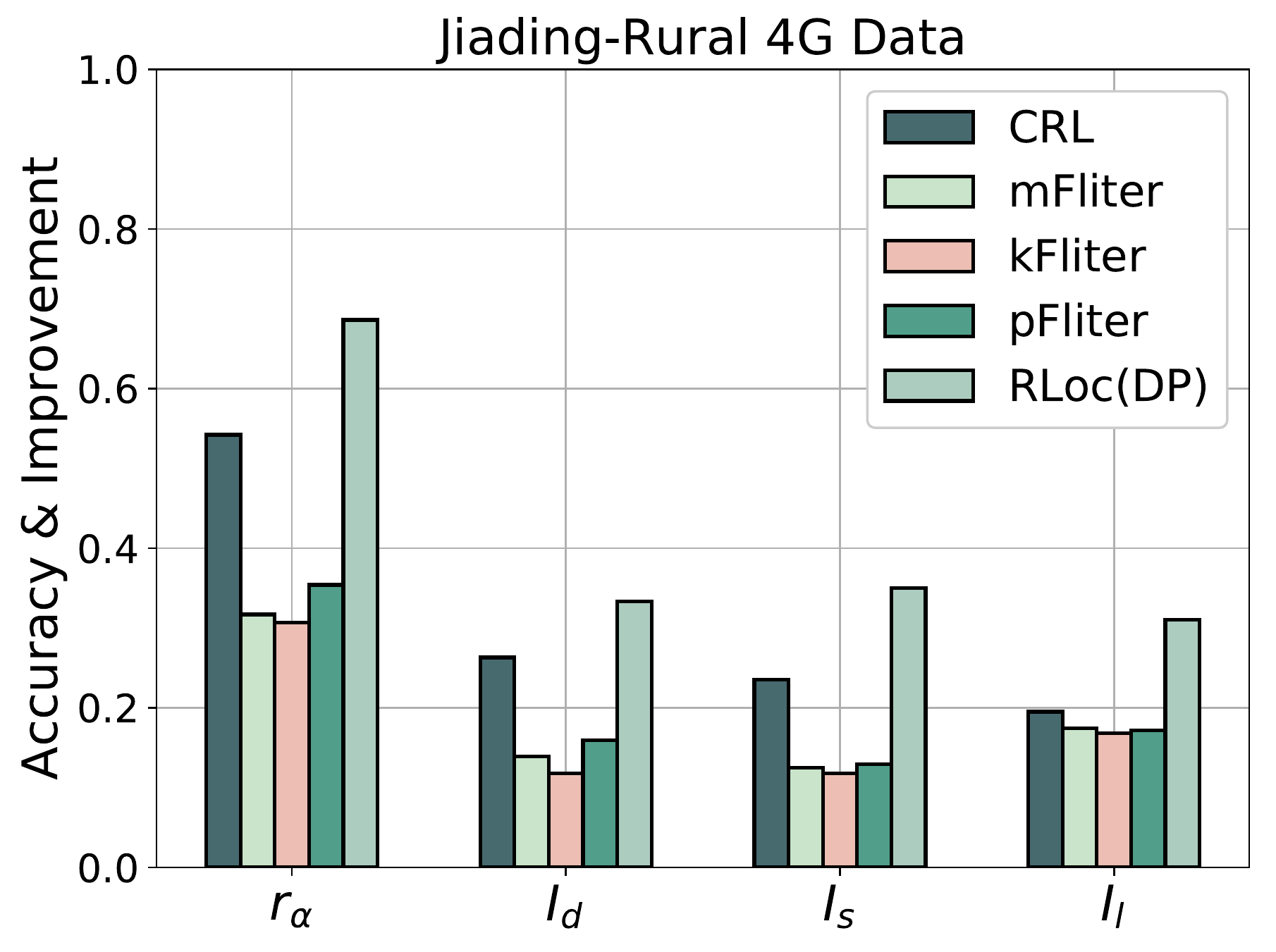}}
					%\centerline{(a)}
				\end{center}
           % \caption{Evaluation of Repair Algorithms.}
		   % \label{exp:repair_alg}%\vspace{-2ex}
			\end{minipage}
            &
            \begin{minipage}[t]{0.47\linewidth}
				\begin{center}
					\centerline{\includegraphics[totalheight=1.1in]{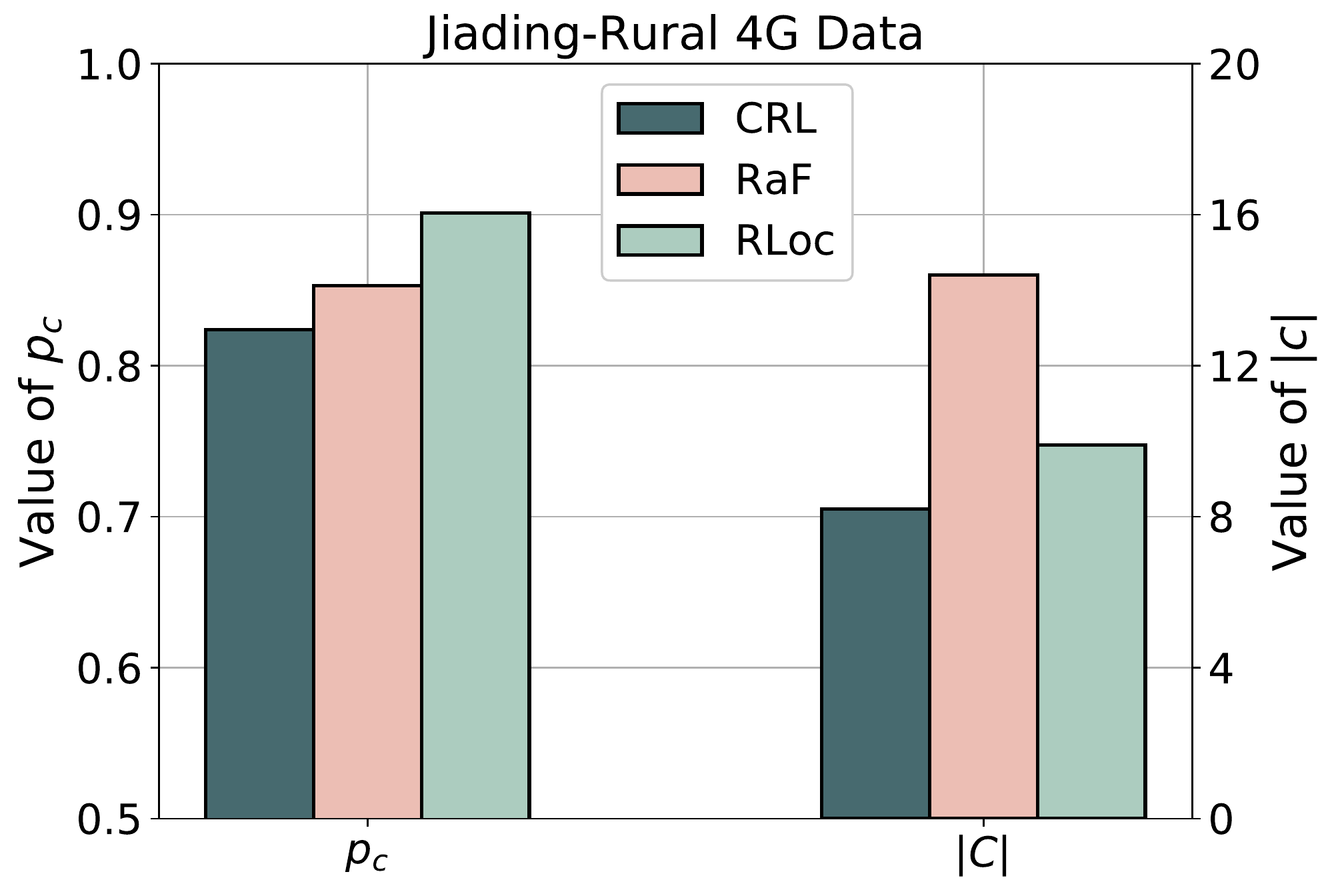}}
					%\centerline{(a)}
				\end{center}
            %    \caption{Effect of Candidate Selection %Criteria in Data Repair.}
		  %\label{exp:select_criteria}%\vspace{-2ex}
			\end{minipage}

        \end{tabular}%\vspace{-2ex}
	\end{center}\vspace{-4ex}
            \caption{Repair Approach: Evaluation of Five Repair Approaches (left) and Effect of Candidate Selection approaches (right).}	\label{sec:repair}
\end{figure}
\subsection{Repair Performance}

In this section, Figure \ref{sec:repair}(a) first evaluates the proposed DP-based repair approach (used by \textsf{RLoc}) against the repair in \textsf{CRL} and three filtering algorithms: mean, Kalman and particle, denoted as \textsf{mFilter}, \textsf{kFilter} and \textsf{pFilter}, respectively. Since both \textsf{CRL} and \textsf{RLoc} require the detection of flawed samples, for fairness, we adopt the same detection approach DA-HMM to select flawed MR samples and then repair these samples by the approaches used by \textsf{CRL} and \textsf{RLoc}. In this figure, the $x$-axis indicates the repair accuracy $r_{\alpha}$ and repair ratios of three localization errors $I_d$, $I_s$ and $I_l$. We find that the DP approach can achieve the highest accuracy $r_{\alpha}$ and greatest repair ratios among all five approaches. It is mainly because the DP approach repairs the entire sequence of flawed MR samples.

Secondly, consider that the candidate set $C$ is the key of a repair algorithm. Thus, we compare our approach against two alternative candidate selection approaches: 1) \textsf{CRL} utilizes a probability Matrix $\mathcal{M}$ to lookup candidates for a given flawed grid, and 2) Random Forest (RaF) classification-based localization model predicts the probability for each possible grid to be the position grid of a testing MR sample. Such a probability can be used to select those top-$k$ grids with the highest probabilities as the candidate grids. In Figure \ref{sec:repair}(b), the left and right $y$-axis plots the repair precision $p_c$ and the number $|C|$ of selected candidates (defined in $\S$ \ref{sec:experimental setting}) of three repair approaches. \textsf{CRL} suffers from the lowest repair precision but selects the smallest candidate grids. Though the Random Forest (RaF) classifier can achieve better result than \textsf{CRL}, but at the cost of the most number of selected candidates (and thus high overhead to prune unneeded candidates). Finally, our approach can achieve the highest precision $p_c$ and the middle amount of selected candidates per flawed sample, 10, is much smaller than the one by RaF. This experiment indicates that our work can lead to the best trade-off between the repair precision and overhead of selecting candidates.

\begin{figure}[th]
	%\hspace{-8ex}
	\begin{center}
		\begin{tabular}{c c}
            \begin{minipage}[t]{0.48\linewidth}
				\begin{center}
					\centerline{\includegraphics[totalheight=1.1in]{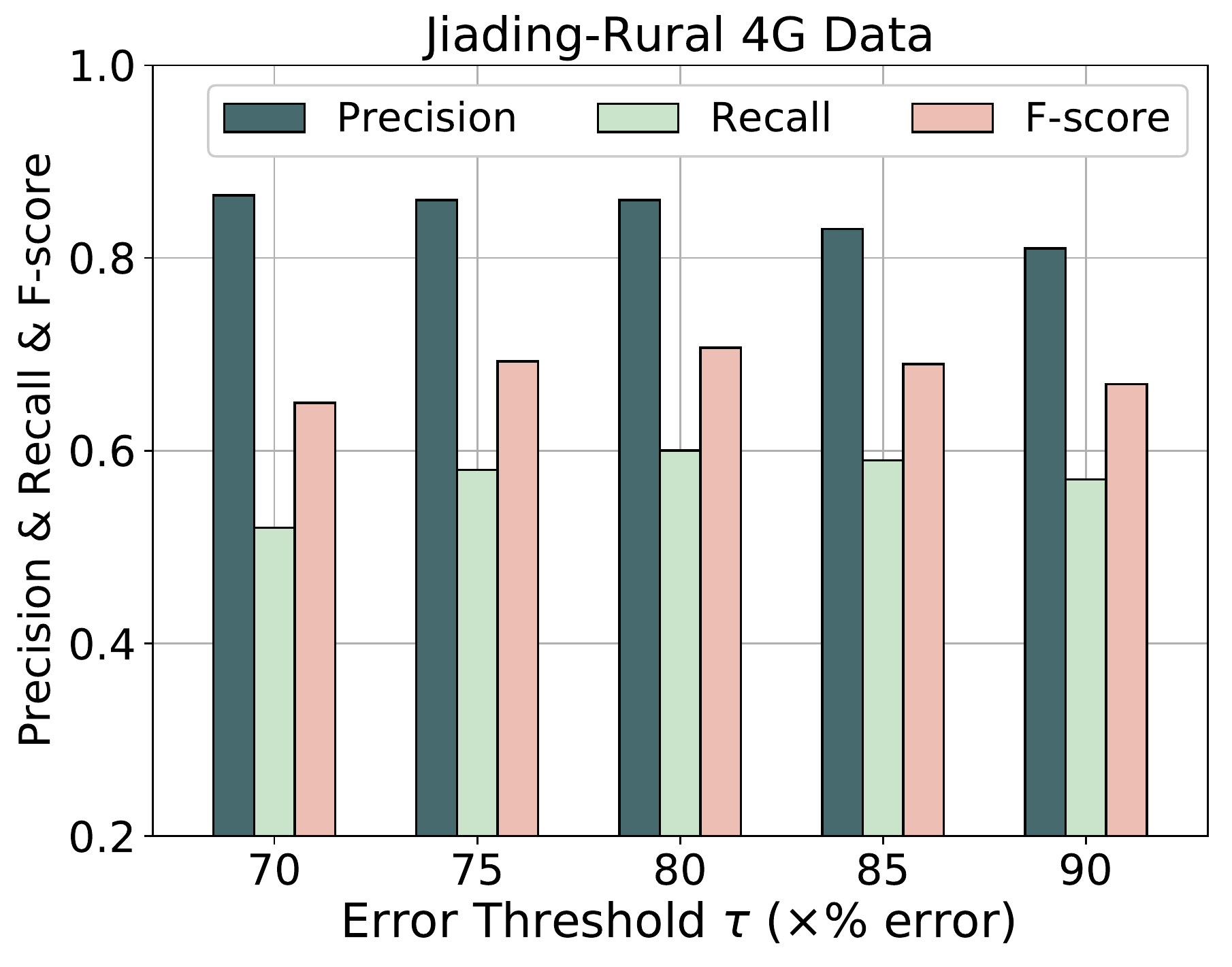}}
					%\centerline{(a)}
				\end{center}
           % \caption{Effect of Error Threshold $\tau$ on %DA-HMM-based Detection.}
		   % \label{exp:error_threshold}%\vspace{-2ex}
			\end{minipage}
            &
            \begin{minipage}[t]{0.48\linewidth}
				\begin{center}
					\centerline{\includegraphics[totalheight=1.1in]{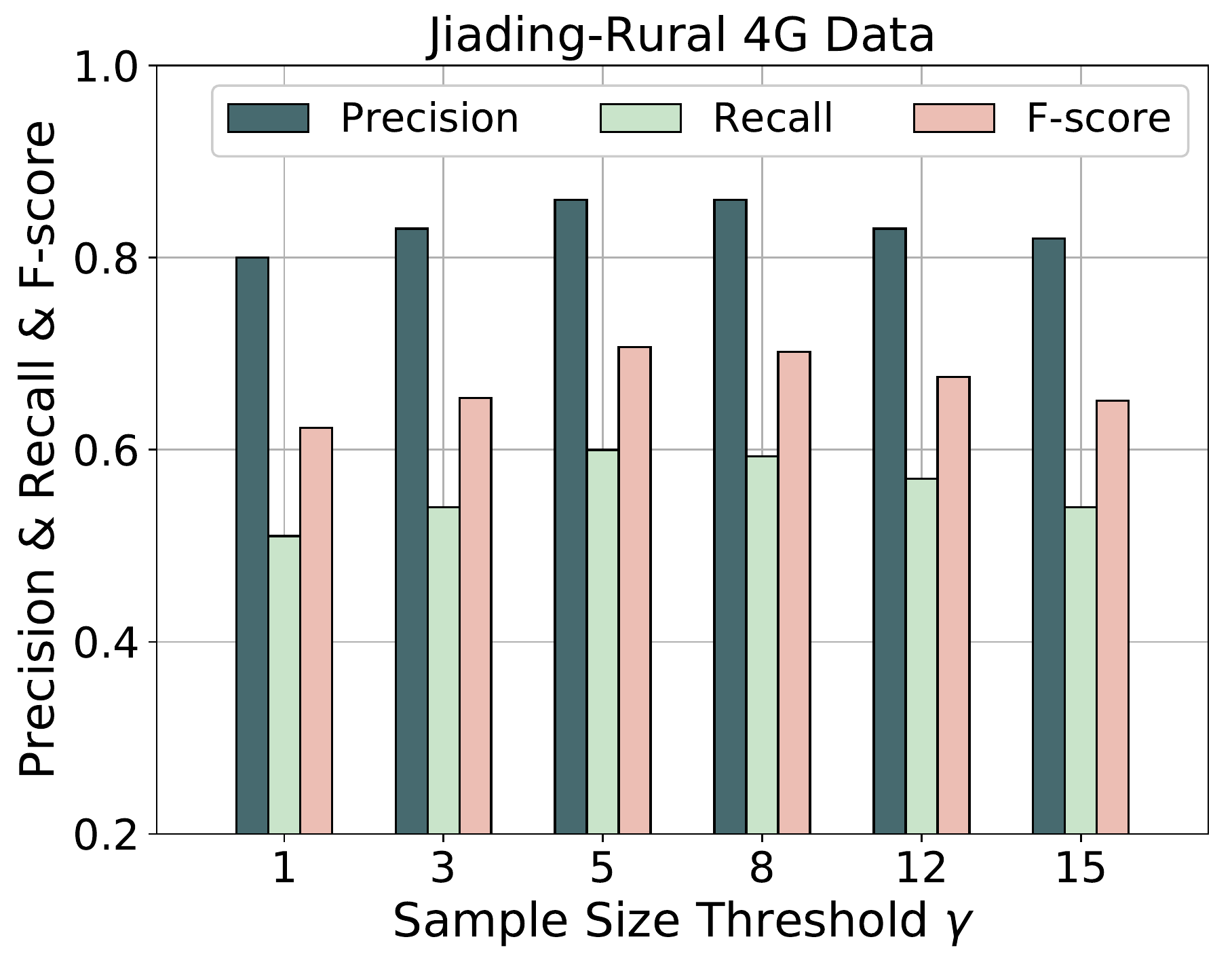}}
					%\centerline{(a)}
				\end{center}
          %  \caption{Effect of Training Sample Size Threshold %$\gamma$ on DA-HMM-based Detection.}
		  %  \label{exp:sample_threshold}%\vspace{-2ex}
			\end{minipage}
        \end{tabular}%\vspace{-2ex}
	\end{center}\vspace{-4ex}
            \caption{Sensitivity Study: Effect of Error Threshold $\tau$ (left) and Sample Size Threshold $\gamma$ (right).}	\label{sec:repair2}
\end{figure}

\subsection{Sensitive Study}\label{sec:sensitive_study}
In this section, we vary the values of several key parameters and study the performance of \textsf{RLoc}. %Note that we give the description of how to determine the value range of key parameters in practice (see Section \ref{sec:experimental setting}), thus the default values of the parameters can be obtained from the results of the following experiments.

%\subsubsection{Effect of Threshold $\tau$ on Detection Algorithm}
Firstly, we are interested in how DA-HMM is sensitive to the threshold $\tau$ which is used to determine the confidence levels to detect flawed samples. Depending upon the errors of a localization model $\mathcal{L}$, we vary the threshold $\tau$ from top-70\% error to top-90\% error (and thus the threshold $\tau$ becomes greater), and measure the performance of the detection algorithm. Figure \ref{sec:repair2}(a) shows the precision, recall and F-score of DA-HMM. When $\tau$ becomes greater, the detection precision drops slightly. Nevertheless, with a larger $\tau$, the recall first grows and later slightly drops. For example, the threshold $\tau=80\%$ error leads to the highest recall. Finally, the F-score unifies precision and recall, and exhibits the similar trend as the recall.

\begin{table}[!hbp]
\caption{Effects of Two Thresholds $\varepsilon$ and $\xi$.}
\label{tab:similarity_thre}
\scriptsize
\centering
\begin{tabular}{|c|c|c|c||c|c|c|c|}
\hline
\multirow{2}{*}{\begin{tabular}[c]{@{}c@{}}Similarity\\ Threshold $\varepsilon$\end{tabular}} & \multicolumn{3}{c|}{Jiading-Rural 4G} & \multirow{2}{*}{\begin{tabular}[c]{@{}c@{}}Similiarty\\ Threshold $\xi$ \end{tabular}} & \multicolumn{3}{c|}{Jiading-Rural 4G} \\ \cline{2-4} \cline{6-8}

& $M_p$ & $M_r$ &$M_f$ & &$r_{\alpha}$ & $p_c$  &$|C|$ \\ \hline
%\hline
0.25&0.833&0.601&0.698 &0.6&0.651&0.907&12.3\cr %\hline
0.50&0.863&0.603&0.710 &0.7&0.686&0.901&9.9\cr %\hline
0.75&0.859&0.582&0.694 &0.8&0.690&0.875&8.2\cr %\hline
1.0 &0.852&0.574&0.686 &1.0&0.694&0.842&6.7\cr \hline
\end{tabular}

\end{table}

%\subsubsection{Effect of Threshold $\gamma$ on Detection Algorithm}
Secondly, we study the effect of the threshold $\gamma$ on our detection algorithm. This threshold $\gamma$ is to determine whether or not the set $|\mathbb{D}_C(v_k^{bs})|$ contains sufficient samples, i.e., $|\mathbb{D}_C(v_k^{bs})|\ge \gamma$, during the estimation of the adaptive emission probability. By varying $\gamma$ from 1 to 15, we evaluate the detection performance of \textsf{DA-HMM}. In Figure \ref{sec:repair2}(b), the precision, recall and F-score of DA-HMM grow until $\gamma=5$ and then degrade slightly. It is mainly because too many observation samples are unnecessary to tune the adaptive probability due to the used sufficient training samples. Thus, we by default set $\gamma$ by 5.

%\subsubsection{Effect of $\varepsilon$ in Detection Algorithm}

Thirdly, we study the effect of $\varepsilon$ in the detection algorithm. Recall that in the DA-HMM detection algorithm, we adopt a similarity threshold $\varepsilon$ to determine whether or not two base station observations are similar. In Table \ref{tab:similarity_thre}, either a too small or too large threshold $\varepsilon$ may not lead to the greatest performance. Here, $\varepsilon=0.5$ helps achieving the best F-score.

%This result indicates that around 5 base stations in $v_k^{bs}$ are contained in $v_k'^{bs}$ as well ($J(v_k^{bs},v_k'^{bs})=\frac{v_k^{bs}\bigcap v_k'^{bs}}{v_k^{bs}\bigcup v_k'^{bs}}=\frac{5}{5+2+2}$, the length of $v_k^{bs}$ cannot exceed seven).

Finally, we are interested in the effect of threshold $\xi$ on our repair algorithm. In the repair algorithm, we use a similarity threshold $\xi$ to determine whether or not a certain spatial grid cell is a candidate. Table \ref{tab:similarity_thre} shows the effect of $\xi$ on the repair performance. Firstly, a greater $\xi$ leads to a smaller number $|C|$ of candidates for each flawed MR sample and higher repair accuracy $a_{\alpha}$. Moreover, the repair precision $p_c$ decreases with a greater $\xi$. For example given $\xi=1.0$, it means that among the up-to 7 base stations in a flawed MR sample, all of them are selected to be the observation set of the candidates. %Note that $\xi=0.9$ has the same effect as $\xi=1$. It is because that if there is only one base station of the flawed MR $r$ is not in the observation set of the candidate grid $g$, then $J'(bs_r, BS_g)\leq \frac{6}{7} = 0.857 <0.9$. Therefore when $\xi=0.9$, only the case where all base stations of the flawed MR are included in the observation of the candidate grid can satisfy $J'(bs_r,BS_g)\geq \xi$. 
Thus, to balance the precision and number $|C|$, we by default set $\xi=0.7$.

\subsection{Discussion}
Telco MR data usually contain privacy sensitive information such as locations and IMSI information of individuals. Privacy preservation techniques can be used to address the privacy issue. For example, we anonymized user identifiers (IMSI) in MR samples. Moreover, we have replaced a real user ID with multiple virtual ones, such that an entire trajectory of this real user could be divided into multiple disjoint sub-trajectories with respect to such virtual users. In this way, we avoid the exposure of an entire trajectory. Nevertheless, the introduced privacy preservation techniques compromise localization accuracy. Our long-term goal is to adopt privacy techniques including differential privacy \cite{DBLP:journals/pvldb/HuYYDCYGZ15} to support privacy-preserving machine learning and accurate data analytics in big Telco MR data.

%% file: 06-conclusion.tex
\section{Conclusion} \label{s:conclusion}

In this paper, we proposed a sequence-based localization framework to detect and repair outlier positions for lower Telco localization errors. First, the detection approach DA-HMM, via a binary confidence level, can overcome the issues of various time intervals of neighbouring MR samples and uneven amount MR samples across base stations. Second, the repair approach leverages a repair graph by incorporating the importance of each candidate and transition between neighbouring candidates to choose a best path with the largest joint probability. The evaluation on three datasets validates that our work greatly outperforms both the single-point-based and traditional sequence-based localization approaches, e.g., those using static HMM models. 

As future work, we continue to explore more advanced machine learning techniques for Telco localization. For example the recent work \cite{8967172} explored transferable knowledge  from training data set to testing data. Such success inspired us to potentially find transferring knowledge between MR samples and GPS locations.

\section*{Acknowledge}
This research has been supported in part by National Natural Science Foundation of China (Grant No. 61972286 and No. 61772371), project 16214817 from the Research Grants Council of Hong Kong and the 5GEAR project and FIT project from the Academy of Finland. We also would like to thank anonymous reviewers for their valuable comments.